\documentclass[reprint,superscriptaddress,floatfix,showpacs,balancelastpage,aps,pra]{revtex4-1}

\usepackage{color}
\usepackage{graphicx}
\usepackage{subfigure}
\usepackage{amsmath}
\usepackage{amssymb}

\usepackage{array}
\usepackage{multirow}

\usepackage[table]{xcolor}
\usepackage[breaklinks]{hyperref}

\renewcommand{\k}[1]{\ensuremath\left|#1\right\rangle}

\newcommand{\irt}{\frac{1}{\sqrt{2}}}

\def\a{\hat a}
\def\ad{\hat a^\dagger}

 \newcommand{\ket}[1]{\ensuremath{\left|#1\right\rangle}}
 \newcommand{\Cre}[0]{\ensuremath{\hat{a}^{\dagger}}}
 \newcommand{\Ann}[0]{\ensuremath{\hat{a}}}
 \newcommand{\OP}[1]{\ensuremath{\hat{#1}}}
 \newcommand{\Sig}[1]{\ensuremath{\hat\sigma_\mathrm{#1}}}
 \newcommand{\Gee}[2]{\ensuremath{g^\mathrm{#1}_{#2}}}
 \newcommand{\eff}[1]{\ensuremath{#1_{\mathrm{eff}}}}

\newlength{\bmgstdfigwidth}
\newlength{\bmgstdfigwidthshorter}
\begin{document}

\title{Cavity QED Photons for Quantum Information Processing}

\author{Moteb M.\ Alqahtani}
\affiliation{Department of Physics and Astronomy,
  University of Sussex, Falmer, Brighton, BN1 9QH,
  United Kingdom}
\author{Mark S.\ Everitt}
\affiliation{National Institute of Informatics, 2-1-2 Hitotsubashi, Chiyoda-ku, Tokyo 101-8430, Japan}
\author{Barry M.\ Garraway}
\affiliation{Department of Physics and Astronomy,
  University of Sussex, Falmer, Brighton, BN1 9QH,
  United Kingdom}

\date{\today.}

\begin{abstract}

  Based on a multimode multilevel Jaynes-Cummings model and multiphoton
resonance theory, a set of universal two- and three-qubit gates, namely the
i\textsc{swap} and the Fredkin gates, has been realized where dual-rail qubits are
encoded in cavities. In this way the information has been stored in cavities
and the off-resonant atomic levels have been eliminated by the semi-classical
theory of an effective two-level Hamiltonian. A further semi-classical model,
namely the spin-$J$ model, has been introduced so that a complete population
inversion for levels of interest has been achieved and periodic multilevel
multiphoton models have been performed. The combination of the two
semi-classical models has been employed to address two-level, three-level,
four-level, and even five-level configurations. The impact of decoherence
processes on the fidelity of the i\textsc{swap} and the Fredkin gates has been studied.   

\end{abstract}

\date{\today}
\maketitle


\section{Photonic logic}

Photonic systems are an attractive choice for quantum information processing
because photons form a natural interface with optical telecommunications and
existing telecoms technology. However, making this choice raises a number of
issues which have to be addressed. It becomes important to be able to
generate single photons, a challenge, but there are now various schemes available
\cite{Lounis, Yamamoto, Matthias}.  A key issue, however, concerns how to process
single photons as qubits and this is the issue addressed in this paper along
with the fact that single photons are very \emph{fragile}. On the other hand
some aspects of photonic computation are very straightforward. For example
the wires of quantum circuits can be represented by optical paths and the
delay of a photonic qubit is a simple way of achieving changes of phase.

In the following we will use the optical language of photons but in the
discussion of the experimental realisation of the models of this work note
that we will also consider microwave photons.

A simple way to realise photonic qubits would be if the presence of a photon
signifies a state $\k{1}$ and the absence of the photon indicates the state
$\k{0}$. To achieve an entangling quantum gate we could try to arrange for
these qubits to interact through the action of a beam-splitter:
\begin{eqnarray}
  \label{eq:map-Q-SBS}
    \k{00} &\longrightarrow&  \k{00} \nonumber\\
    \k{01} &\longrightarrow&  \irt(\k{01} + i\k{10})  \nonumber\\
    \k{10} &\longrightarrow&  \irt(\k{10} + i\k{01}) \nonumber\\
    \k{11} &\longrightarrow&  \irt(\k{20} +  \k{02}) .
\end{eqnarray}
However, this unfortunately produces states like $\k{20}$ which are not
qubit encodings in the original simple scheme.
An alternative approach would be to try a non-linear cross-Kerr interaction
\cite{Imoto, Yamamoto2} where
$$
  H_{x}  = \chi_x \ad_1 \a_1 \ad_2 \a_2 .
$$
Then the time evolution $\exp(-i H_{x} t /\hbar)$, acting on $\k{00},
\k{01}$ and $ \k{10}$, results in no change to the state, and the input
state $ \k{11}$ is modified by a phase factor $\exp(-i \chi_x t /\hbar )$.
If $ \chi_x t /\hbar = \pi$, the CZ gate (controlled-$Z$ gate), could be
realised where: \index{CZ gate}\index{gate!CZ}
\begin{eqnarray}
  \label{eq:map-Q-CZ}
    \k{00} &\longrightarrow&  \k{00} \nonumber\\
    \k{01} &\longrightarrow&  \k{01} \nonumber\\
    \k{10} &\longrightarrow&  \k{10} \nonumber\\
    \k{11} &\longrightarrow& -\k{11} .
\end{eqnarray}
This can be represented by the table
\begin{equation}
  \label{eq:mat-Q-CZ}
  \mbox{CZ} \equiv \left[ \begin{array}{rrrr}%
1& & & \\
 &1& & \\
 & &1& \\
 & & &-1\\
\end{array}\right]  .
\end{equation}
Here, and in the following we will assume that such tables are expressed in
the standard qubit basis $\{\k{00}, \k{01}, \k{10}, \k{11}\}$.  The CZ-gate
is a universal quantum gate when combined with single qubit rotations.
However, for the Hamiltonian $H_{x} = \chi_x \ad_1 \a_1 \ad_2 \a_2$ the
non-linearity is too weak to achieve this simple gate in this way
\cite{Schmidt, Kok2}.

In the linear optical quantum computing vision \cite{Kok} non-linear optics
can be avoided in qubit processing by including photon detection in the
computation process. Measurement projection is an interruptive and hence
non-linear process which can be configured for quantum information
processing and which includes heralding to indicate the successful arrival
of photons through the processor. In achieving this the KLM scheme (Knill,
Laflame, Milburn \cite{KLM}) was a breakthrough. Here the undesirable output
states $\k{02}$ and $\k{20}$ are avoided by the use of a special non-linear
sign gate indicated by NLS. The NLS gate has the special property that for
the general input state $ \alpha\k{0} + \beta\k{1} + \gamma\k{2}
\longrightarrow \alpha\k{0} + \beta\k{1} - \gamma\k{2}$ so that there is a
sign change for the amplitude of the $\k2$ state. This means that this gate
has no effect on the three basic input states $\{\k{00}, \k{01}, \k{10}\}$
which work well with a beam-splitter, and thus the gate acts as a
Mach-Zehnder interferometer. For the $\k{11}$ input, however, the change of
sign from the NLS components ensures that the output is a sign flipped
$\k{11}$ state.  However, the NLS gate has a maximum theoretical probability
of 1/4 for success \cite{Rudolph}. This can start to be an issue if very
large numbers of gate operations are required.

For these reasons we will move away from flying photonic qubits in the
following and examine the possibilities for stationary photonic qubits in a
cavity.  The advantage is that we already know we can realise an extremely
strong non-linear interaction between photonic qubits, i.e.\ through a
multimode Jaynes-Cummings model (JCM). We already know in the case of the
single-mode JCM that the interaction can be so strong that the system
exhibits Rabi oscillations \cite{Jaynes}. However, a critical issue to
examine is the decay of the system, both through cavity decay and through
atomic decay (cavity modified spontaneous emission).

In the following we consider some multimode, multi-$\Lambda$ cavity QED
systems for quantum information processing. We start with a simple two-mode,
single $\Lambda$ system in section~\ref{sec:Larson}. After examining this
system's weaknesses we introduce a four-mode double-$\Lambda$ system in
section~\ref{sec:MultiphotonLogic}. Section~\ref{sec:FourModesVariants}
examines the dynamics of the four mode resonance with and without intermediate resonant
states: intermediate resonant states can achieve a speed-up. This study includes an examination of
decoherence effects. Section~\ref{sec:ThreeAndOneQubitGates} examines some
three-qubit gates and single qubit rotations. The paper concludes in
section~\ref{sec:Conclude} and several Appendices follow with details of
some of the calculations of effective Hamiltonians for these multi-state,
multimode systems.

\section{A two-mode $\Lambda$-system}
\label{sec:Larson}

\begin{figure}\centering
  
   \includegraphics[width=\bmgstdfigwidthshorter]{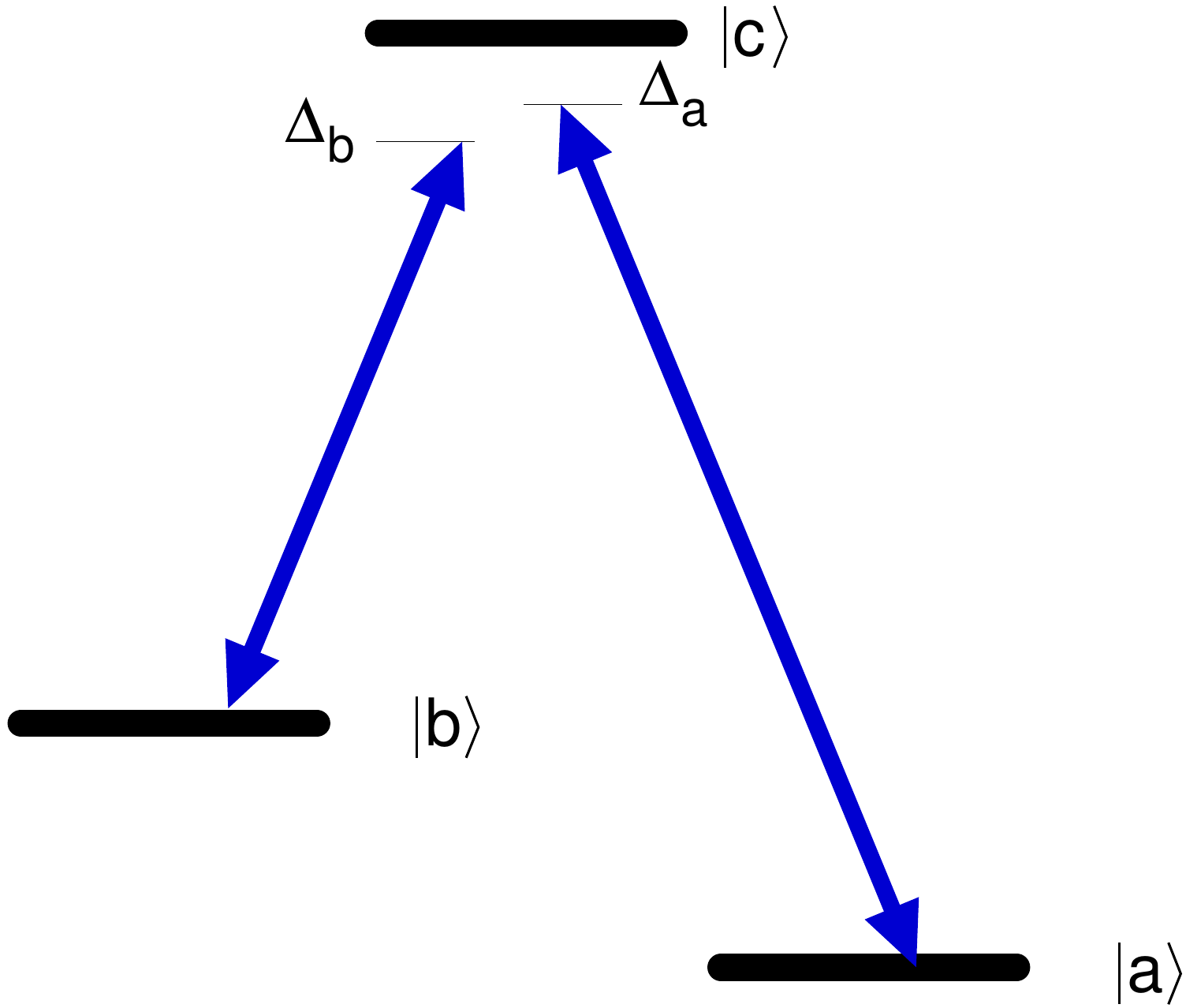}

  \caption{A three-level atom in $\Lambda$-configuration, with lower levels
    $|a\rangle$, $|b\rangle$ and upper level $|c\rangle$, interacts with two
    cavity modes $\omega_1$ and $\omega_2$. The states $|b\rangle$ and
    $|a\rangle$ couple to $|c\rangle$ through a dipole interaction with modes
    one and two respectively.} 

\label{fig:Larson}
\end{figure}

In Ref.\ \cite{Barry2}, a system involving two EM modes interacting with a
three level atomic $\Lambda$-system (see Fig.~\ref{fig:Larson}) has been
examined.
%
%
%
The Hamiltonian for this system, in the Rotating Wave Approximation (RWA), is then given by
\begin{align}\label{eq:LarsonHamileff}
 H_I = &-\Delta_b\OP{n}_1 + \Gee{cb}{1}\left(\Ann_1\Sig{bc} + \Cre_1\Sig{cb}\right)
\nonumber       \\
&-\Delta_a \OP{n}_2 + \Gee{ac}{2}\left(\Ann_2\Sig{ac} +
         \Cre_2\Sig{ca}\right)\ ,
\end{align}
where $\hat{a}_{1}$ and $\hat{a}_{2}$ are the boson operators for the two
modes, the atomic operators have the form $\sigma_{\mu \nu}\equiv
|\mu\rangle \langle \nu|$, and $g_{1,2}$ are the atom-field coupling
constants. The definitions of the system detunings are
$\Delta_a=(\omega_c-\omega_{a})-\omega_2$ and
$\Delta_b=(\omega_c-\omega_b)-\omega_1$.

We can consider the limit for a two-photon process when the detuning
$\Delta_a = \Delta_b \rightarrow \Delta$. In this limit we can perform a
standard adiabatic elimination \cite{Gerry} to realise an effective
Hamiltonian
\begin{equation}
  H=g(\hat a_2^{\dagger}\hat a_1\hat \sigma^-+\hat a_2 \hat
  a_1^{\dagger}\sigma^+)\ ,
\label{Larsoneffham}
\end{equation}
where we have introduced operators that effectively make direct transitions
between the two ground states, i.e.\ $\hat\sigma^+=|b\rangle\langle a|$,
$\hat\sigma^-=|a\rangle\langle b|$.  The adiabatic elimination results in an
effective coupling
\begin{equation}
 g=g_{ac}g_{bc}/\Delta \ .\label{eq:Larsongeff}
\end{equation}
This system makes transitions between the two ground states if it is
possible to extract a photon from one of the modes and transfer it to the
other one.
We can quickly find a general solution to the Hamiltonian
(\ref{Larsoneffham}) which is reminiscent of the Jaynes-Cummings model
itself, but with two-index terms in the Rabi frequency because of the two
modes \cite{Barry2}: 
\begin{align*}
|\Psi(t)\rangle \simeq &
\sum_{n,m}C_n^{(1)}C_m^{(2)}\left\{c_a\left[\cos(gt\sqrt{(n+1)m})|n,m,a\rangle\right.\right.
\\ &  \left.-i\sin(gt\sqrt{(n+1)m})|n+1,m-1,b\rangle\right]
\\ &
+ c_b \left[\cos(gt\sqrt{(m+1)n})|n,m,b\rangle\right. \\ &
\left.\left.-i\sin(gt\sqrt{(m+1)n})|n-1,m+1,a\rangle\right]\right\}.
\end{align*}
We have considered the potential for logic operations, but note that for
some obvious input states we have mappings where state $\k{1,1,a}\rightarrow
\k{1,1,a},\k{2,0,b}$ at frequency $g\sqrt{2}$, and state
$\k{0,1,a}\rightarrow \k{0,1,a},\k{1,0,b}$ at frequency $g$. These
frequencies are non-commensurate which makes it awkward to eliminate
$\k{2,0}$ and obtain a standard gate.
However, for higher photon numbers the state $\k{3,3,a}\rightarrow
\k{3,3,a},\k{4,2,b}$ at frequency $2g\sqrt{3}$ and the state
$\k{0,3,a}\rightarrow \k{0,3,a},\k{1,4,b}$ at frequency $g\sqrt{3}$, which
are commensurate frequencies.
Thus, when $gt\sqrt{3}=\pi$ the atom becomes an ancilla and 
\begin{equation}\label{phgate}
\begin{array}{lcl}
|0,0,a\rangle\rightarrow|0,0,a\rangle & &
|0,0,b\rangle\rightarrow|0,0,b\rangle \\ 
|0,3,a\rangle\rightarrow-|0,3,a\rangle & &
|0,3,b\rangle\rightarrow|0,3,b\rangle \\ 
|3,0,a\rangle\rightarrow|3,0,a\rangle & &
|3,0,b\rangle\rightarrow-|3,0,b\rangle \\ 
|3,3,a\rangle\rightarrow|3,3,a\rangle & &
|3,3,b\rangle\rightarrow|3,3,b\rangle.
\end{array}
\end{equation}
In this way we can realise the CZ gate for photons (Eq.~(\ref{eq:mat-Q-CZ}))
provided we identify the logical qubits:
\begin{align*}
  \text{mode 1:}&&\k0 &\rightarrow \text{`1',} &\k3 &\rightarrow \text{`0'}\\
  \text{mode 2:}&&\k0 &\rightarrow \text{`0',} &\k3 &\rightarrow \text{`1'}
\,.
\end{align*}
However, this immediately raises a number of issues relating to the use of a
higher Fock state $\k3$, such as decoherence, and how to initialise the
qubits efficiently and how to perform single qubit rotations? For these
reasons we take a different approach.

\section{Multiphoton logic}\label{sec:MultiphotonLogic}

\begin{figure}
  \centering
  \includegraphics[width=\bmgstdfigwidthshorter]{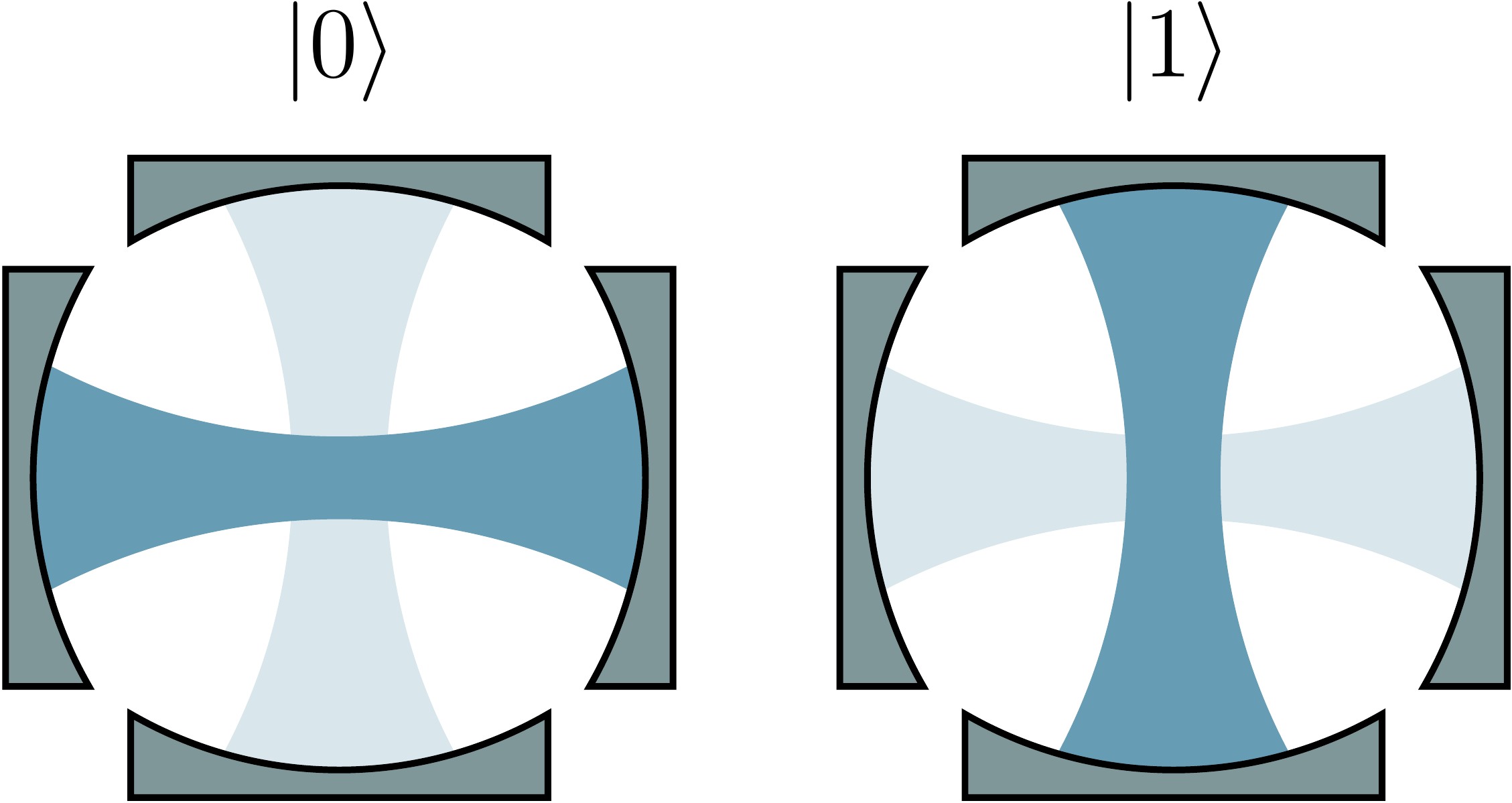}

  \caption{Schematic showing the relationship of logical qubits to cavity mode
    excitations (indicated as a darker colour). The case of two cavity modes
  is illustrated.}
\label{fig:mode-picture}
\end{figure}

To bring about solutions, or near solutions to these problems mentioned at the
end of the previous section we will make an adaptation to the scheme
presented there. This adaptation is to encode our cavity qubits
as dual-mode cavity qubits \cite{Barry1}. The more familiar term ``dual-rail''
seems inappropriate here as the qubits are not flying and there is no
rail. The dual mode qubits are formulated, for example, as shown in
Table~\ref{tab:qubit-encoding}. 
\begin{table}[h]
  \centering
  \begin{tabular}{ccc}
 EM Modes && Logical qubit\\
 \hline
 $\ket{1}\ket{0}$ &$\mapsto$& $\ket{1}$\\
 $\ket{0}\ket{1}$ &$\mapsto$& $\ket{0}$\\
 \end{tabular}

 \caption{Dual-mode qubit coding of the type used in this section.}
\label{tab:qubit-encoding}
\end{table}
In this approach a qubit state is always encoded with a single excitation, as
in the dual rail approach \cite{Yamamoto2, Yamamoto3, Kok} and this
ensures that if a cavity decay process takes place, we have: $\ket{1}\ket{0}
\rightarrow \ket{0}\ket{0}$, and then the result is not a valid qubit in this
encoding. The difference between the two logical qubit states is where the
excitation of the two modes is located, as indicated in
Fig.~\ref{fig:mode-picture}.

\begin{figure}

 \includegraphics[width=0.35\textwidth]{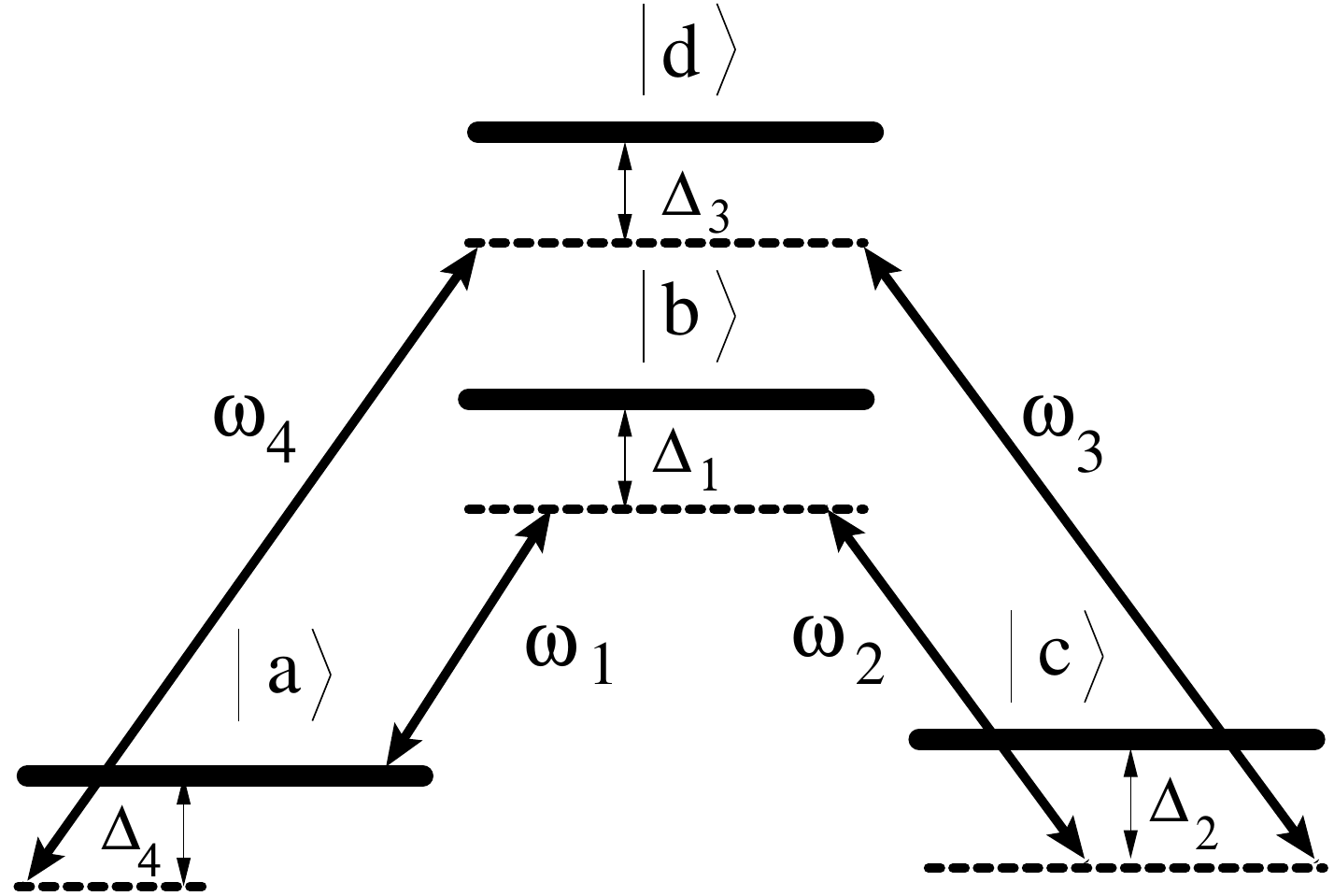}
 
  \caption{Scheme of two-qubit i\textsc{swap} gate, where $\omega_i$ (with $i=1,2,3,4$) are
    modes of the four-mode high $Q$ cavity, and $|a\rangle$, $|b\rangle$,
    $|c\rangle$, and $|d\rangle$ are four energy levels of the atom.}
\label{fig:FourLevels}
\end{figure}

A consequence of using the dual-mode approach is that in order to have a
two-qubit gate we need to have an interaction of an atom with \emph{four}
cavity modes, which is quite demanding (as we shall see), and it is this
which we investigate below. As in the simple two-mode system of section
\ref{sec:Larson} and Hamiltonian~(\ref{eq:LarsonHamileff}), we will achieve
this by having a sequence of off-resonant interactions with a multilevel
atom (Fig.~\ref{fig:FourLevels}), and, as in section \ref{sec:Larson}, that
atom will be an ancilla which will not be entangled with the final result of
a gate operation. The aim of the off-resonant interactions is to reduce
spontaneous emission from the ancilla atom. In order that we do not extract
``information'' from the system via the ancilla atom it is important for the
sequence of atomic transitions to start and end on the atomic state $\k{a}$
(see Fig.~\ref{fig:FourLevels}).  Multiphoton resonance will also require
the final detuning, $\Delta_4$ to ensure resonance. This will mean
$\Delta_4\approx 0$, but it has to be optimised for level shifts as we see
below.


The full Hamiltonian of the four-level system is then \cite{Jaynes, Shore2}:
\begin{eqnarray} 
H&=&\sum \limits_{i=a,b,c,d}\ \omega_{i}\ \hat{\sigma}_{ii}+\sum
\limits_{j=1}^4 \ \omega_j\ \hat{a}^{\dag}_j \hat{a}_j \\
\nonumber&&+ [\Gee{ab}{1}\ \hat{a}_1\ \hat{\sigma}^{ba}+\Gee{bc}{2}\
\hat{\sigma}^{cb}\ \hat{a}^{\dag}_2+\Gee{cd}{3} \hat{a}_3\
\hat{\sigma}^{dc}+\Gee{da}{4}\ \hat{\sigma}^{ad}\ \hat{a}^{\dag}_4\\
\nonumber&&+\rm{h.c.}], \label{eq:iSWAP Hamiltonian}
\end{eqnarray} 
where the first two terms represent the non-coupling Hamiltonians, and the
remaining terms describe the atom-field interaction Hamiltonian. 
%
The four key initial states of the system and their mappings can be
summarised as 
\begin{align} \label{eeq:key-states1}
\nonumber \text{physical}&\qquad\text{logical}\\
\nonumber \ket{a,\ 0110}&\mapsto\ket{a,\ 00}\\
          \ket{a,\ 0101}&\mapsto\ket{a,\ 01}\\
\nonumber \ket{a,\ 1010}&\mapsto\ket{a,\ 10}\\
\nonumber \ket{a,\ 1001}&\mapsto\ket{a,\ 11} \ .
\end{align}
\begin{figure}[h]\centerline{
\subfigure[]{\includegraphics[width=0.2\textwidth]{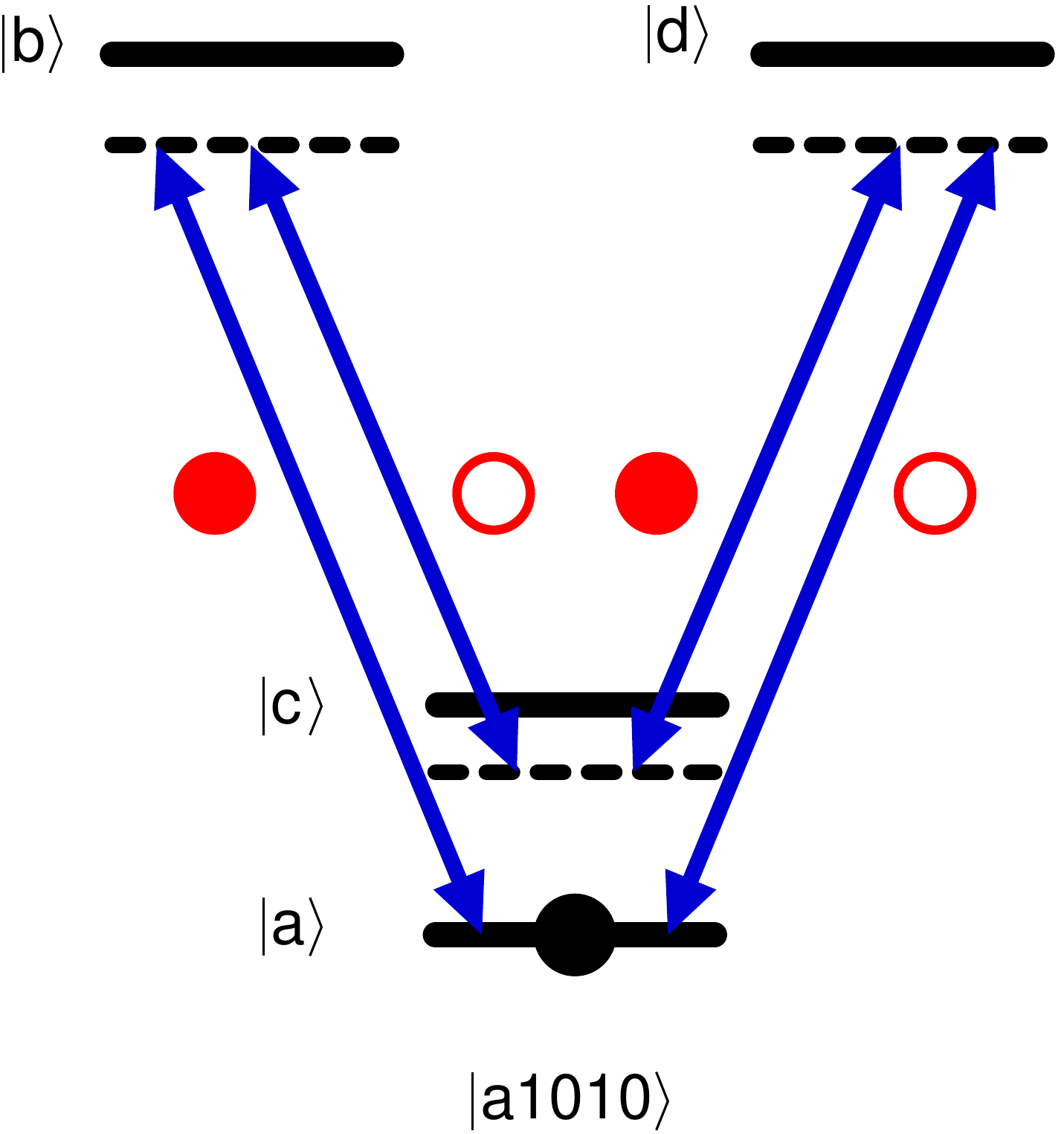}}
\subfigure[]{\includegraphics[width=0.2\textwidth]{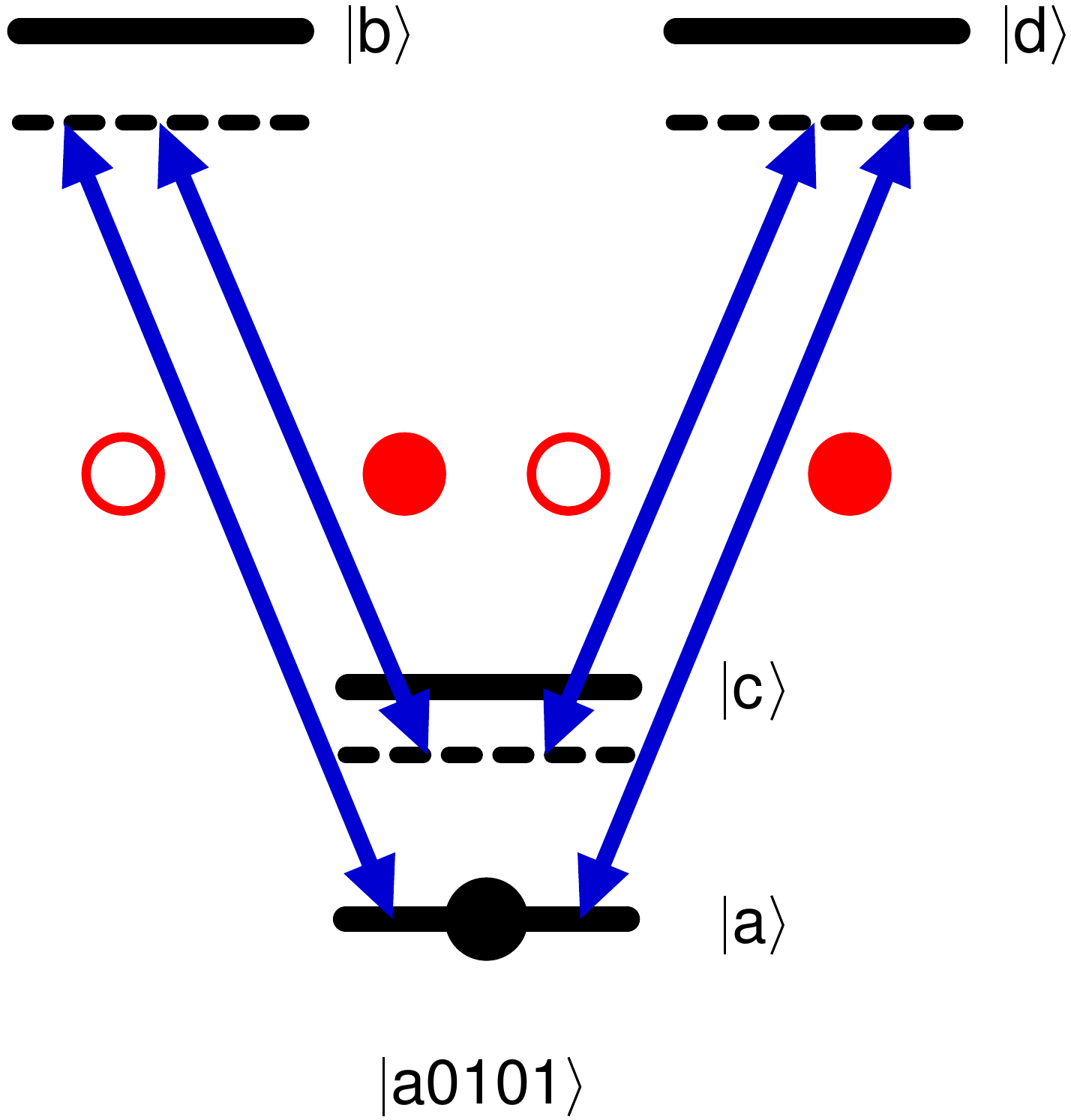}}}
\centerline{
\subfigure[]{\includegraphics[width=0.2\textwidth]{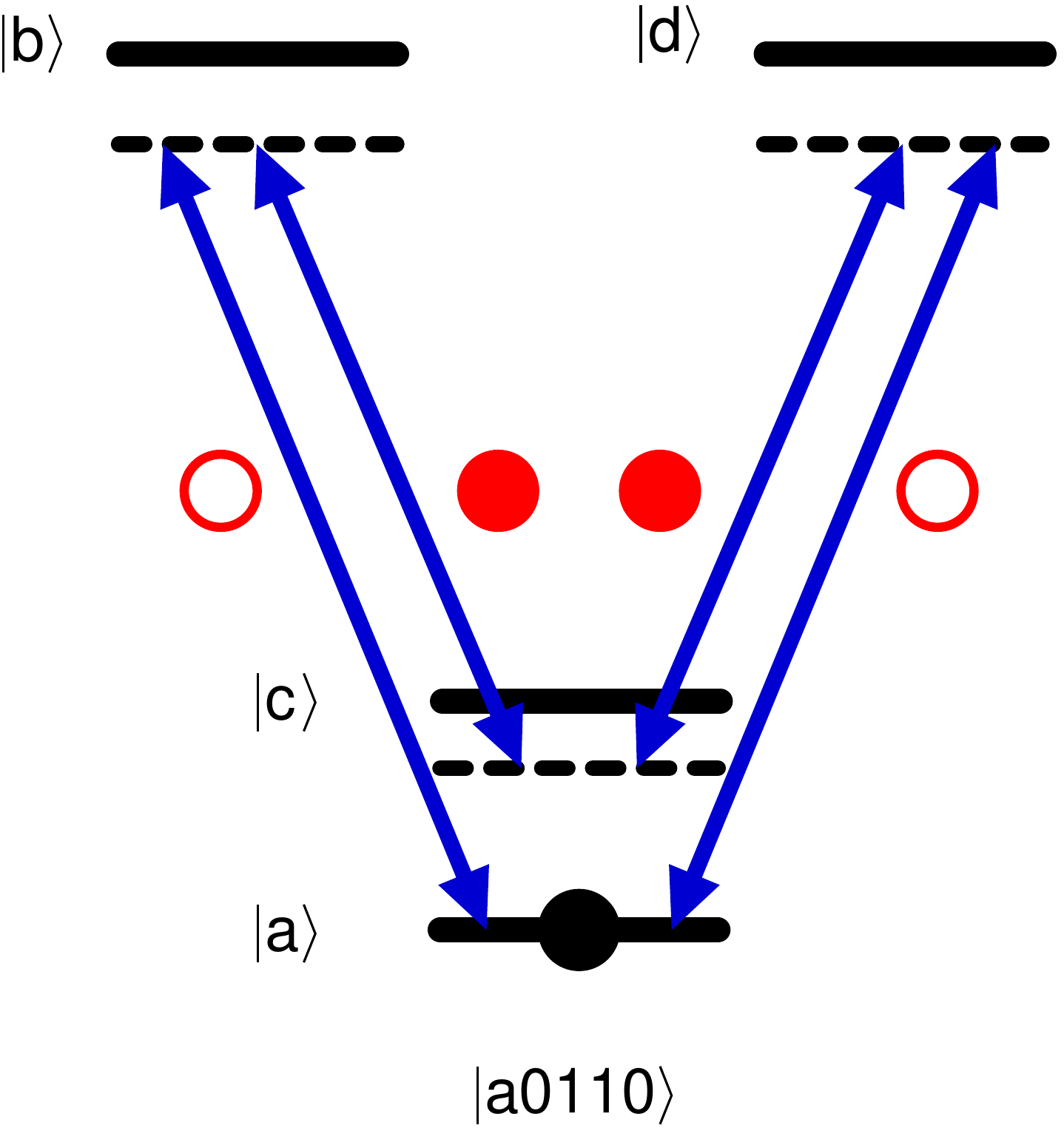}}
\subfigure[]{\includegraphics[width=0.2\textwidth]{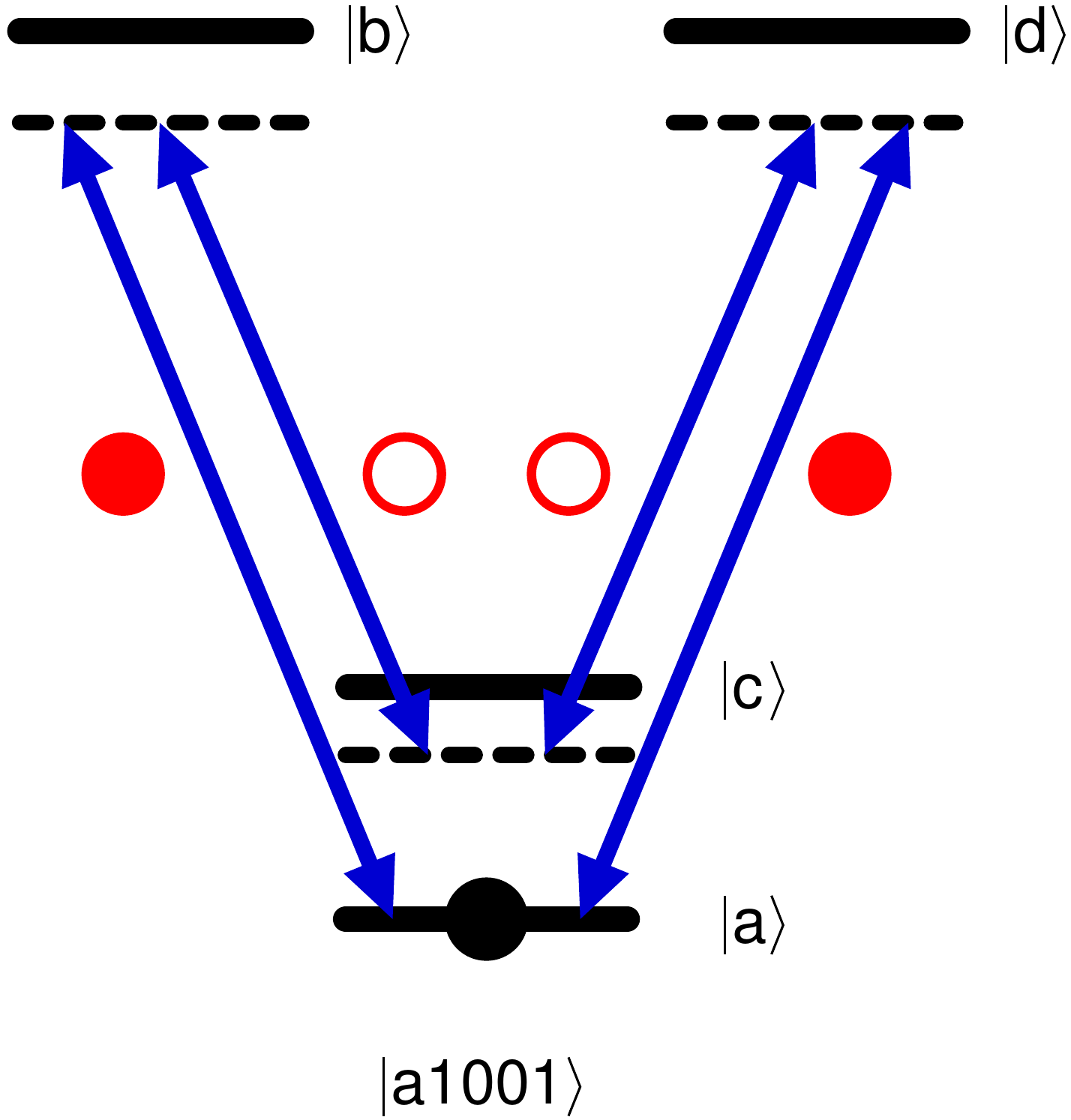}}}
\caption{The four key initial states of the system with possible
  interactions. Empty red circles represent empty cavity modes and filled
  red circles represent cavity modes with a single excitation. The filled
  black circles indicate where the atomic population is located in this
  simplified analysis of the gate process. Starting from atomic state $\k a$ we can see that
(a) and (b) can progress, but (c) and (d) are blocked.} \label{fig:4level-blocked}
\end{figure}
In the case of the initial state $\ket{a,\ 1010}$, the system (atom+field) is
governed by the Hamiltonian $H^{\prime}$ which can be expressed, in the matrix
representation and with $\ket{a,\ 1010}$ to be the zero-point energy, as

\begin{eqnarray}
H^{\prime}=\left(\begin{array}{ccccc}0&\Gee{ab}{1}&0&0&0\\
                                    \Gee{ab}{1}&\Delta_1&\Gee{bc}{2}&0&0\\
                                    0&\Gee{bc}{2}&\Delta_2&\Gee{cd}{3}&0\\
                                    0&0&\Gee{cd}{3}&\Delta_3&\Gee{da}{4}\\
                                    0&0&0&\Gee{da}{4}&\Delta_4 \end{array}\right)\
                                ,\label{eq:iSWAP_int_Ham} 
\end{eqnarray}
in the basis states $\{|a\ 1010\rangle,|b\
0010\rangle,|c\ 0110\rangle,|d\ 0100\rangle,\\ |a\ 0101\rangle \}$.


Should we start with different logical qubits, i.e. different
arrangements of the excited modes, the multiphoton process is
blocked in the case of $|a\ 0110\rangle$. In Fig.~\ref{fig:4level-blocked}(c) 
there is no cavity photon available in the first and last modes to raise the
atomic state from $\k{a}$. When the initial state is $|a\ 1001\rangle$, on the
other hand, the evolution of the system is governed by the Hamiltonian
$H^{''}$ given as 
\begin{eqnarray}
H^{''}=\left(\begin{array}{ccccc}\Delta_2&\Gee{bc}{2}     &0       &0       &0     \\
                             \Gee{bc}{2}    &\Delta_1&\Gee{ab}{1}     &0       &0       \\
                             0      &\Gee{ab}{1}     &0       &\Gee{da}{4}     &0       \\
                             0      &0       &\Gee{da}{4}     &(\Delta_3-\Delta_4)&\Gee{cd}{3}     \\
                             0      &0       &0       &\Gee{cd}{3}     &(\Delta_2-\Delta_4)
        \end{array} \right)\ ,\label{eq:a1001 Hamiltonian}
\end{eqnarray}
acting in the basis states $\{|c\ 0101\rangle,|b\ 0001\rangle,|a\
1001\rangle,\\ |d\ 1000\rangle,|c\ 1010\rangle\}$.

\begin{figure}

  \includegraphics[width=0.23\textwidth]{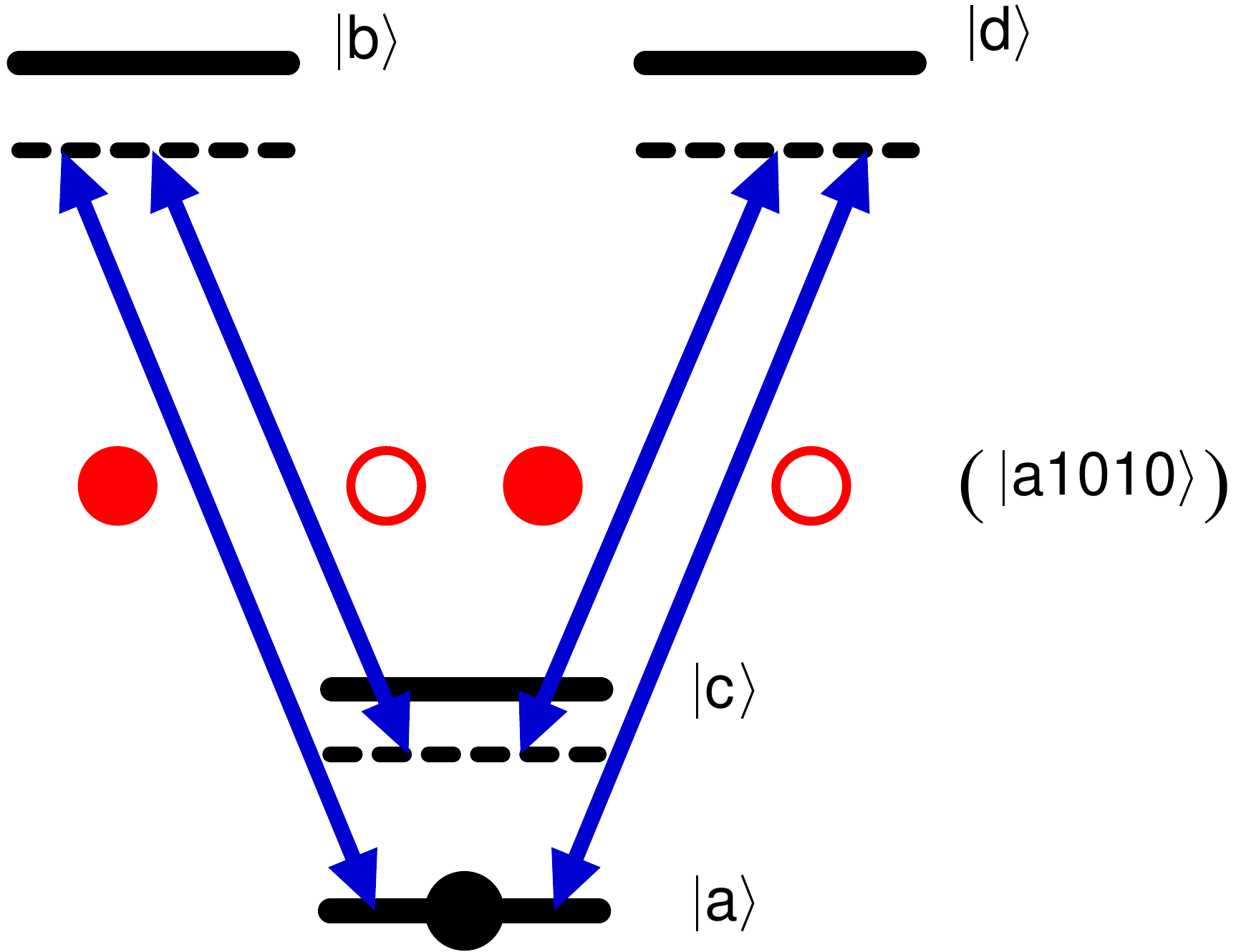}\llap{
 \parbox[b]{3.5in}{(a)\\\rule{0ex}{.8in}}}

  \includegraphics[width=0.23\textwidth]{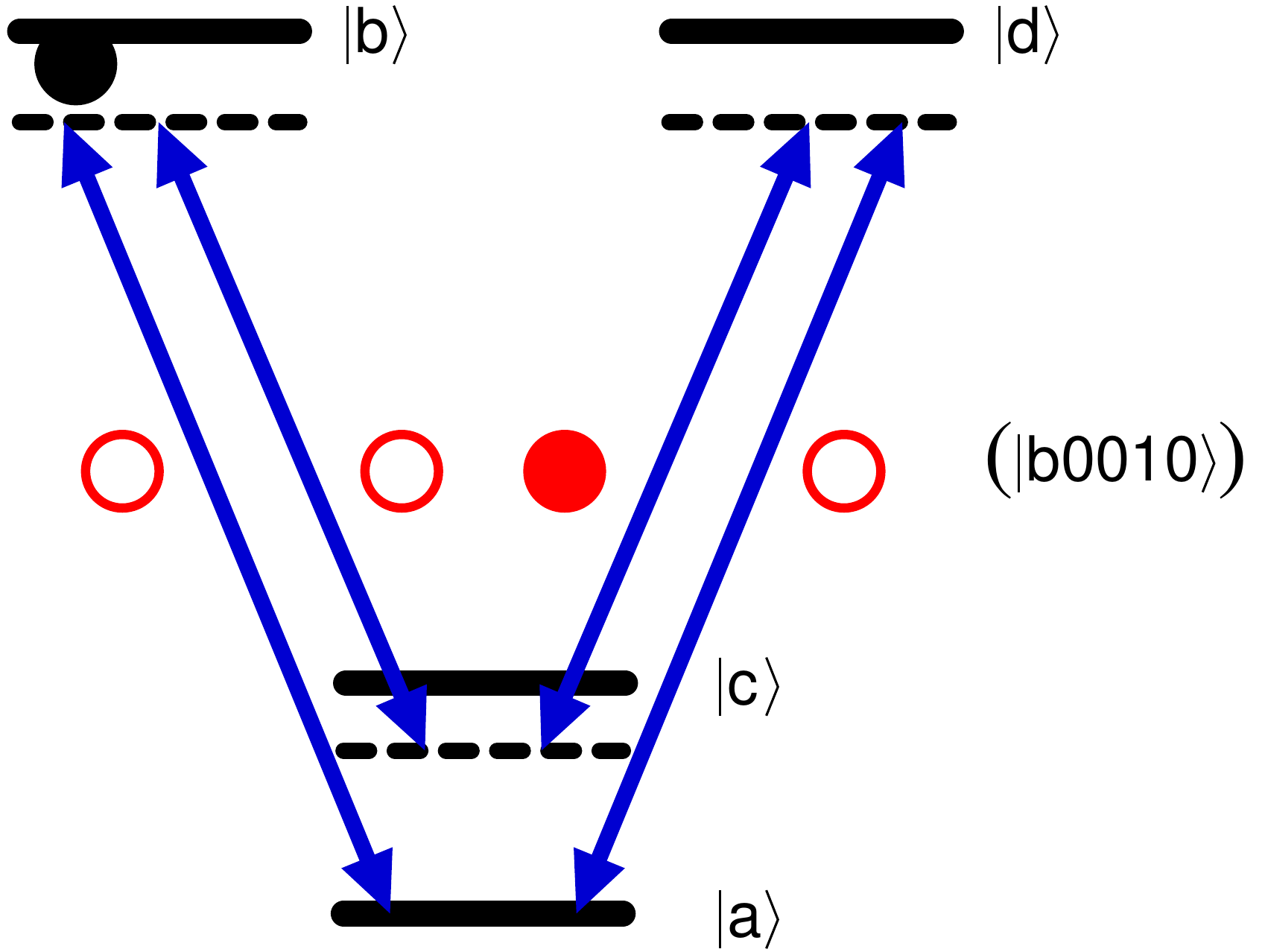}\llap{
 \parbox[b]{3.5in}{(b)\\\rule{0ex}{.7in}}}

  \includegraphics[width=0.23\textwidth]{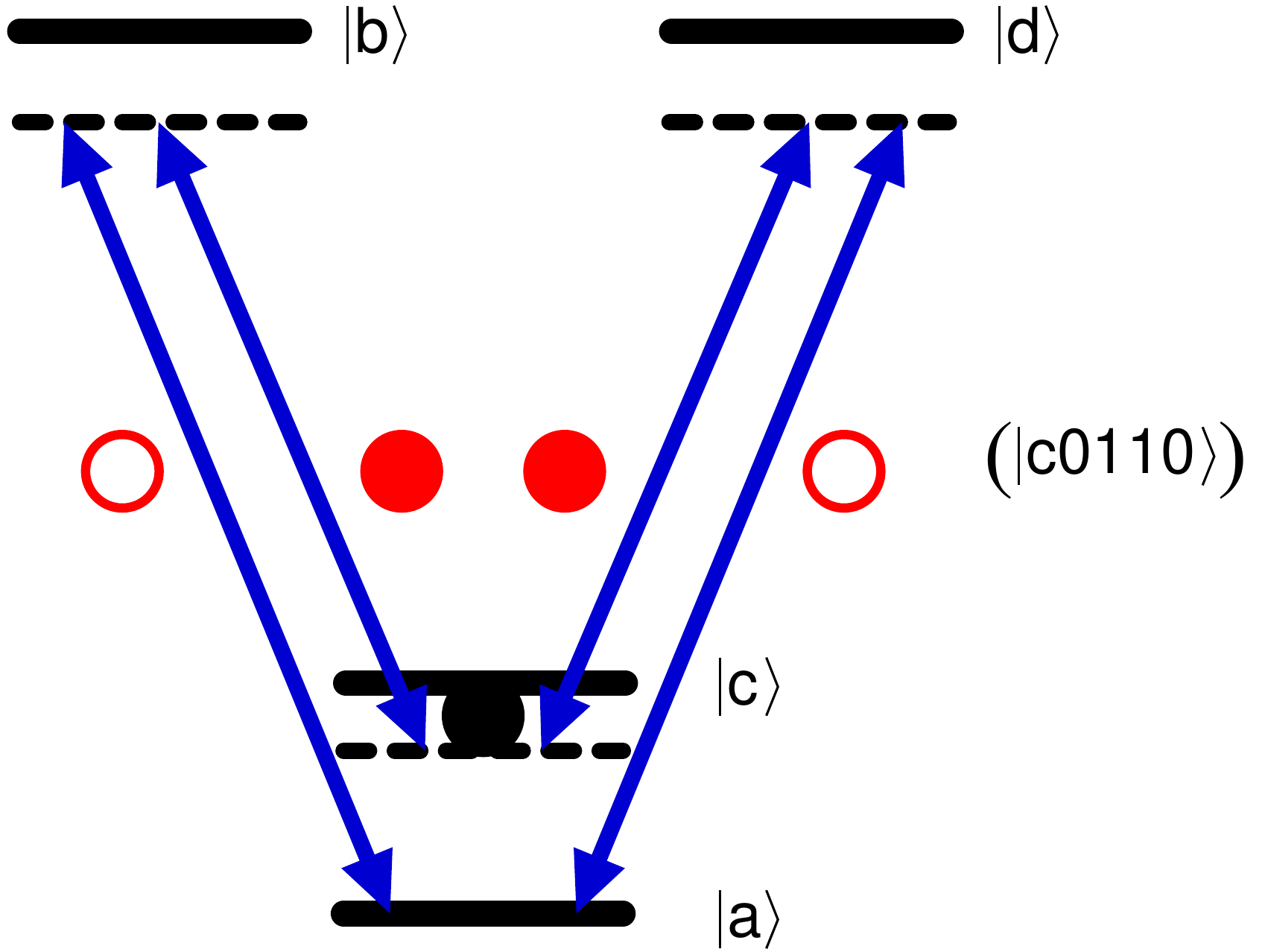}\llap{
 \parbox[b]{3.5in}{(c)\\\rule{0ex}{.6in}}}

  \includegraphics[width=0.23\textwidth]{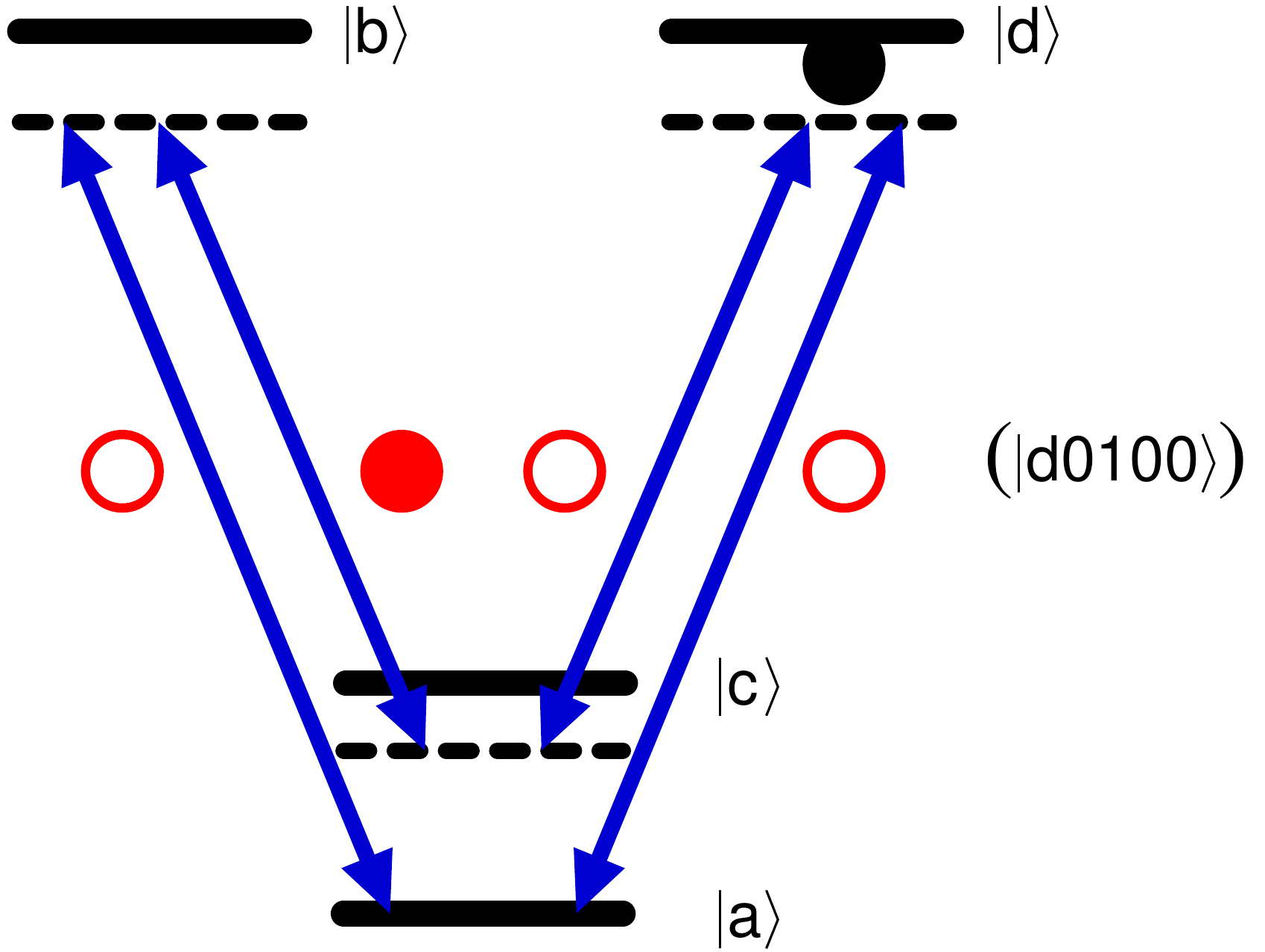}\llap{
 \parbox[b]{3.5in}{(d)\\\rule{0ex}{.5in}}}

  \includegraphics[width=0.23\textwidth]{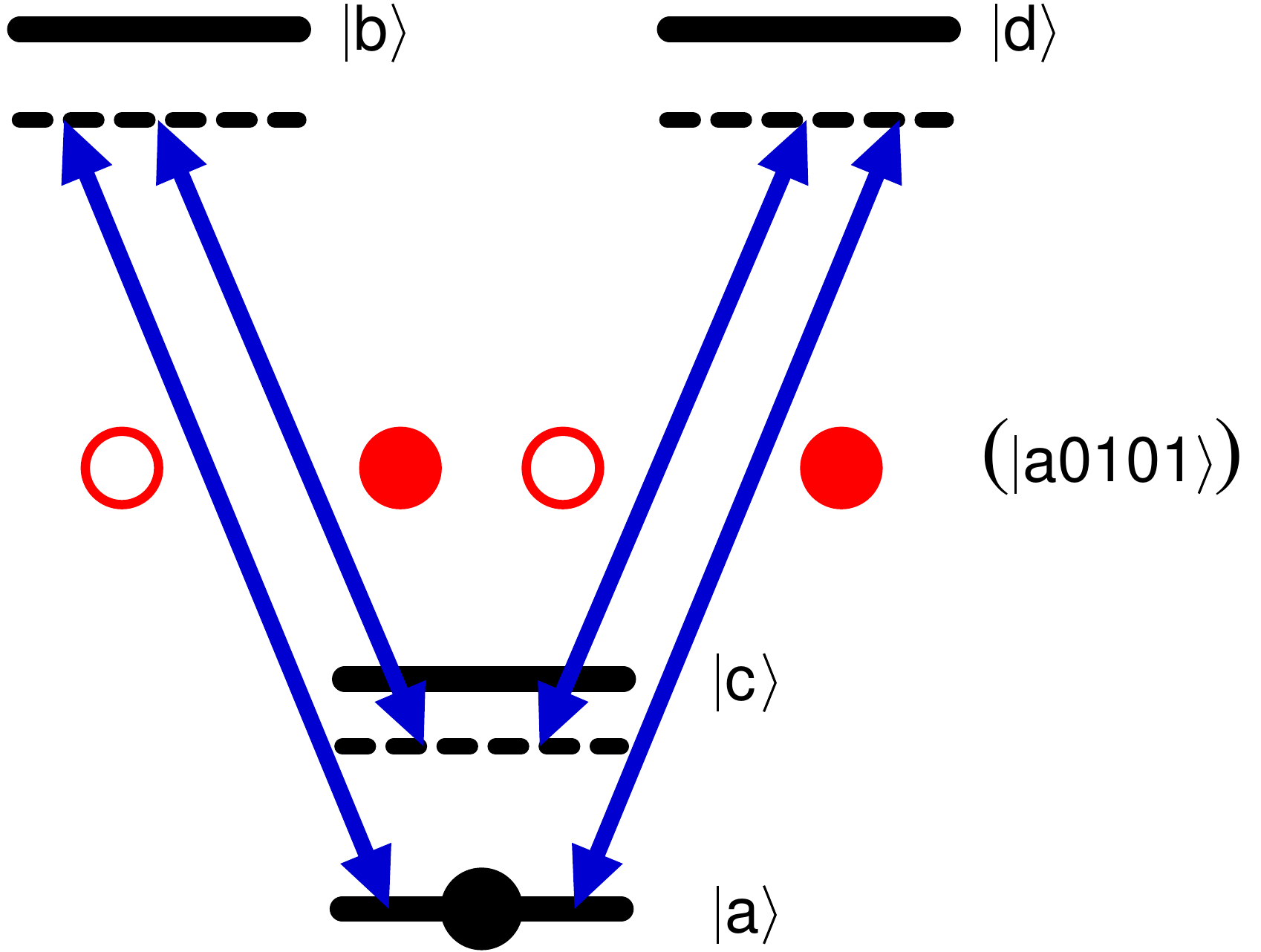}\llap{
 \parbox[b]{3.5in}{(e)\\\rule{0ex}{.4in}}}
  
  \caption{Sequence of steps for the shuffling excitation amongst four cavity
    modes. This shuffling realises the swapping part of a gate operation,
    i.e.\ $|a\ 1010\rangle \leftrightarrow |a\ 0101\rangle$.} 
\label{fig:4level-sequence}
\end{figure}

The way the multiphoton process works to shuffle cavity excitation can be
seen from Fig.~\ref{fig:4level-sequence}. An initial state is set up in
Fig.~\ref{fig:4level-sequence}(a) where the first and third modes are excited
and the second and fourth are not and the atom is in state $\k{a}$. Moving to
Fig.~\ref{fig:4level-sequence}(b) the atom state moves to $\k{b}$ by means of
the absorption of a photon from the first mode. Subsequently this can be
emitted into the second mode, Fig.~\ref{fig:4level-sequence}(c), as the atom
approaches the off-resonant state $\k{c}$.
Figures~\ref{fig:4level-sequence}(d) and \ref{fig:4level-sequence}(e) show how
the atomic state returns to $\k{a}$ by means of a similar process via the
off-resonant level $\k{d}$.

\section{Variants of 4 modes}
\label{sec:FourModesVariants}

\begin{figure}

  \renewcommand{\arraystretch}{2.0}
  \begin{tabular}[t]
{rm{0.2\textwidth}}
     \multicolumn{2}{c}{2 resonant states}\\
    Model(10001)&\includegraphics[width=0.2\textwidth]{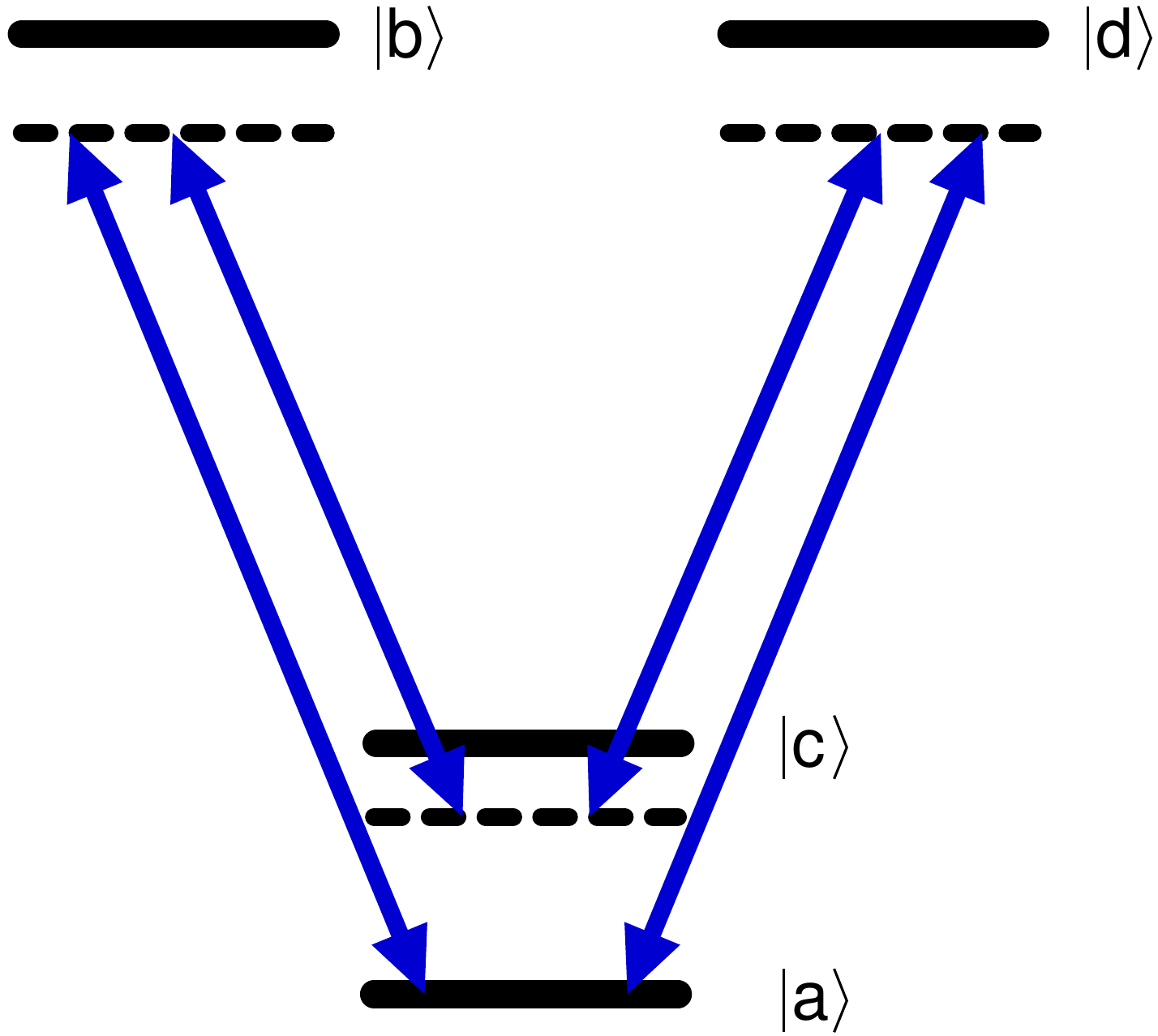}\\
     \multicolumn{2}{c}{3 resonant states}\\
    Model(10101)&\includegraphics[width=0.2\textwidth]{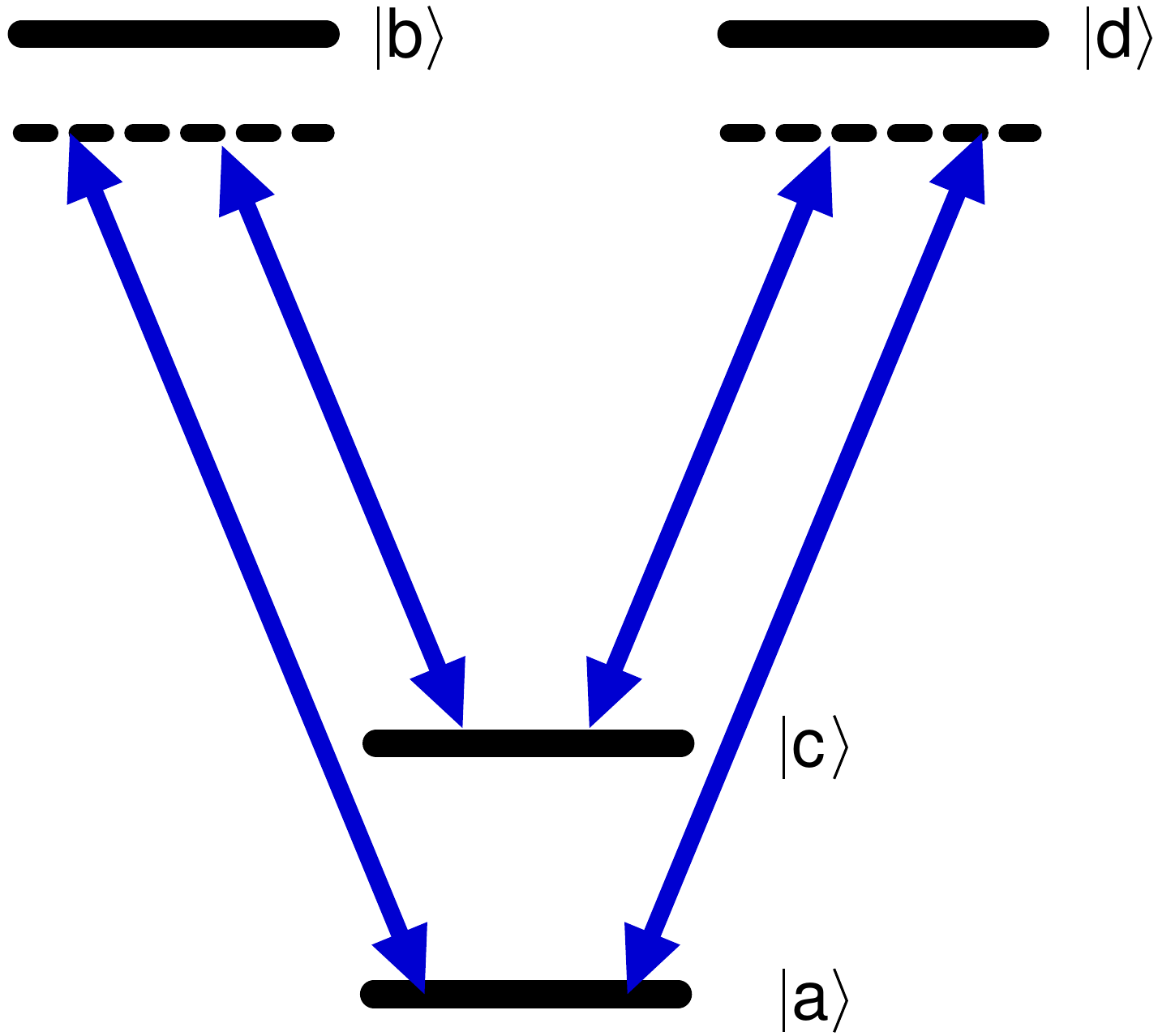}\\
    Model(11001)&\includegraphics[width=0.2\textwidth]{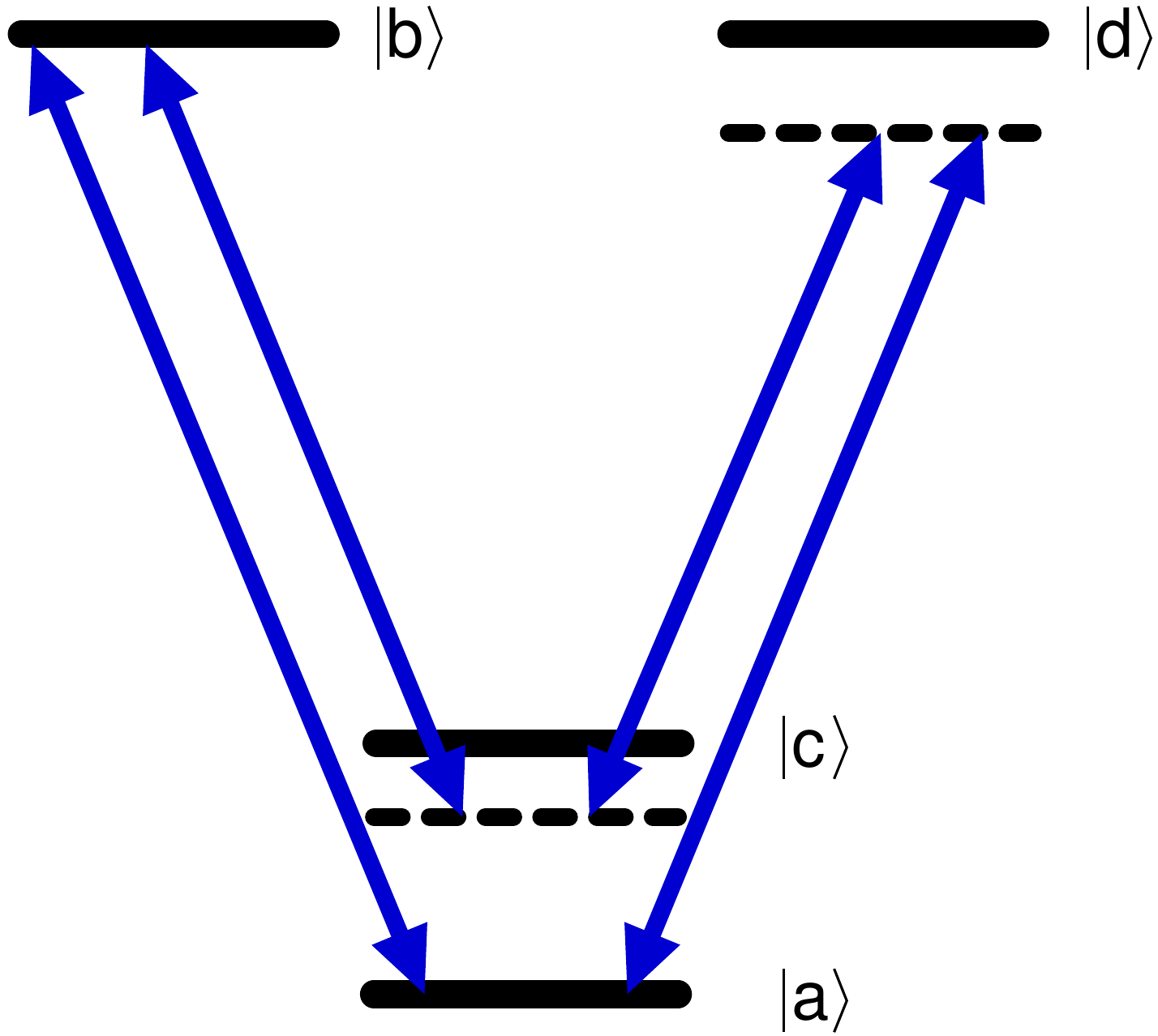}\\

     \multicolumn{2}{c}{4 resonant states}\\
    Model(11011)&\includegraphics[width=0.2\textwidth]{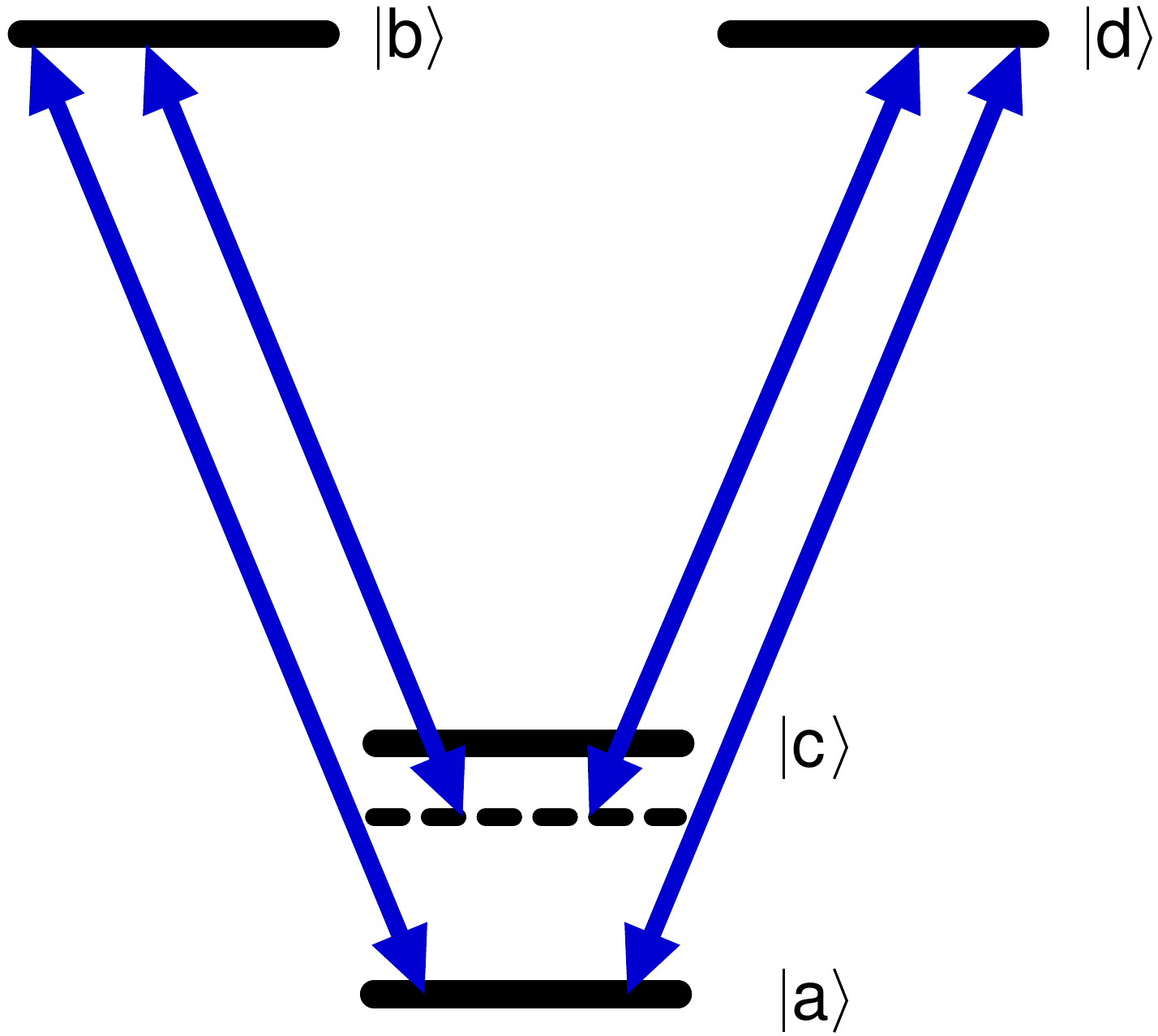}\\
    Model(10111)&\includegraphics[width=0.2\textwidth]{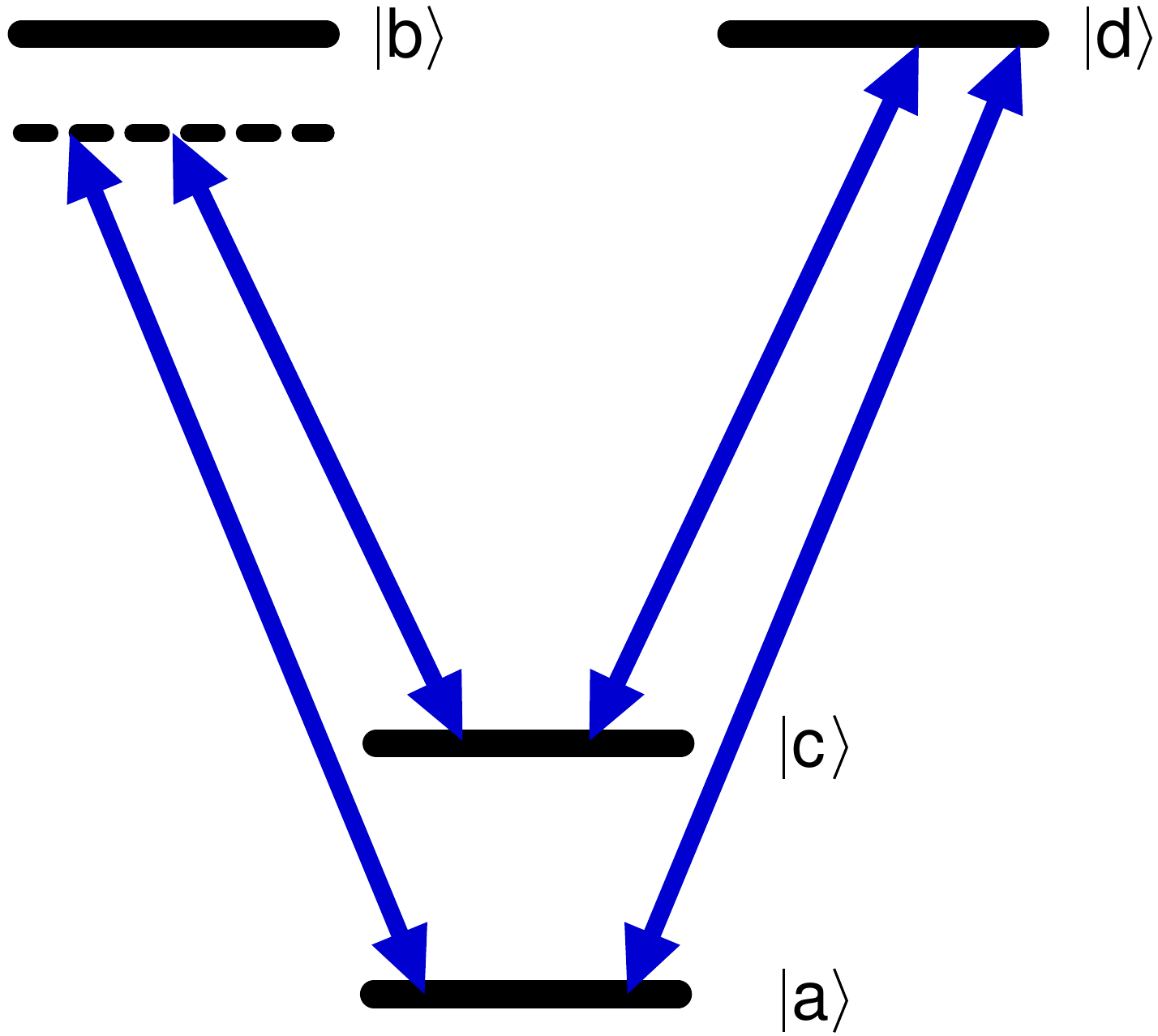}\\
     \multicolumn{2}{c}{5 resonant states}\\
    Model(11111)&\includegraphics[width=0.2\textwidth]{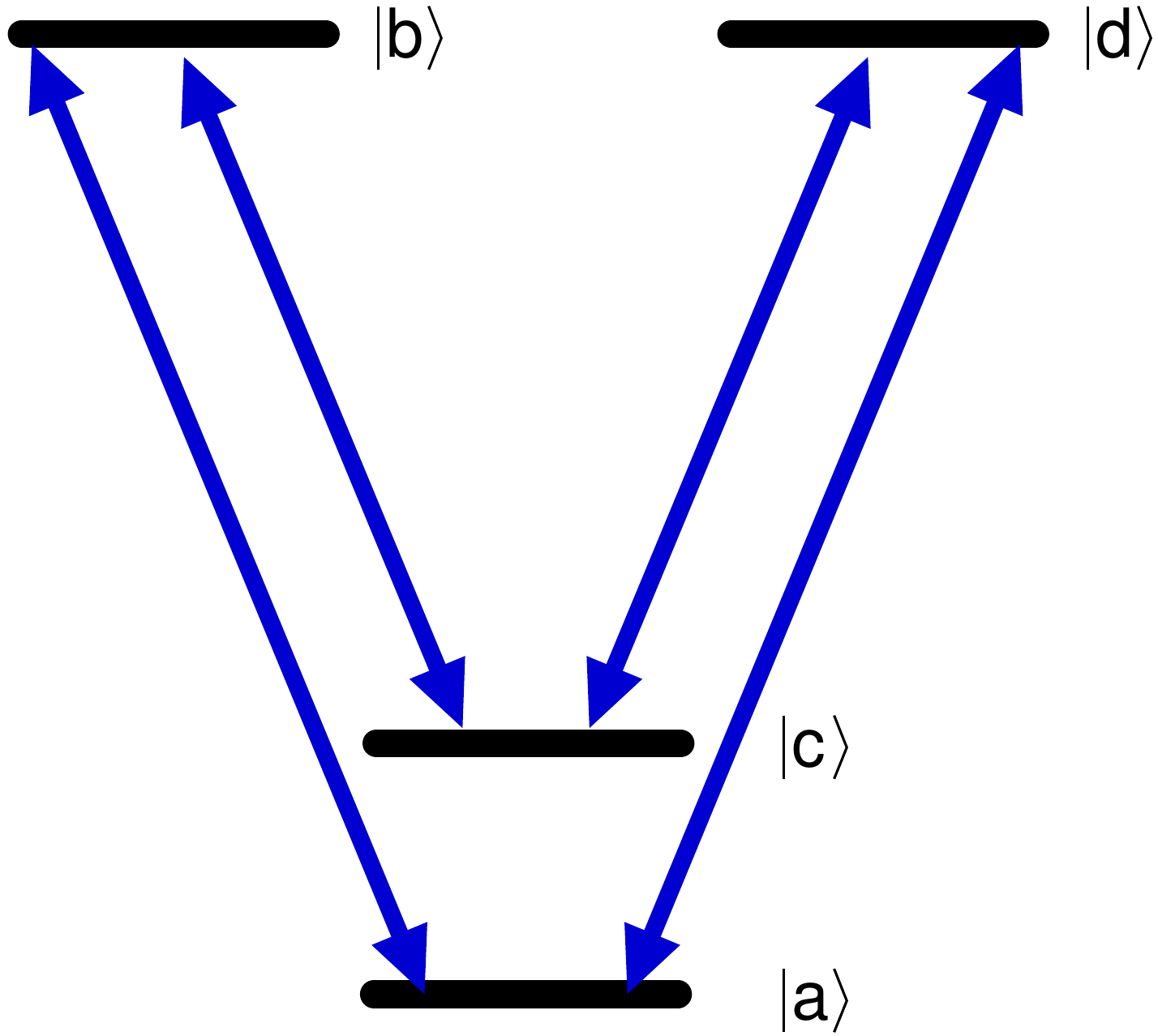}\\
  \end{tabular}

\caption{Linkage schemes for 4 modes N resonant states (with
  N$=1,2,3,4,5$).} \label{fig:Models3and4} 
\end{figure}

In this section we will examine several alternate approaches to carrying out
the swapping part of the gate. Because the time evolution of the i\textsc{swap} gate
exhibited in the following Sec.~\ref{sec:iSWAP2level} is extremely slow, we
can achieve a speed-up by allowing some of the intermediate states in
Fig.~\ref{fig:FourLevels} to become resonant. However there are a wide range
of possibilities which are shown in Fig.~ \ref{fig:Models3and4}.  Our choice
of system needs to be informed by which configurations can keep the qubit
state $|a 11\rangle$ in its initial state at the appropriate interaction
time.

To easily identify the different model systems we will use a binary type
notation to indicate which levels in the sequence of states are to be
resonant, and which are not. Thus the Model(10001) in
Fig.~\ref{fig:Models3and4} has only two resonant states in the chain, the
first and last ones, and will be treated in Eq.~(\ref{eq:Heff-iSWAP-slow})
If we took a fully resonant model, where every transition is resonant, the
model would be Model(11111) also shown in Fig.~\ref{fig:Models3and4}. In the
next section we will focus on Model(11001), as one of the cases with three
resonant levels, i.e.\ it has two intermediate resonant levels.

\subsection{An effective two-level system} 
\label{sec:iSWAP2level}

For the initial state $|a\ 1010\rangle$, the effective wave-function of the
system can be expressed as:
\begin{eqnarray}\label{eq:fourmodebasis}
\nonumber
|\Psi(t)\rangle&=&c_1 |a1010\rangle +c_2|b0010\rangle+c_3|c0110\rangle
\\
&&+ c_4|d0100\rangle+c_5  |a0101\rangle \ .
\end{eqnarray}
which is a superposition of the key states used for single photon swapping. 

In Appendix~\ref{sec:TheoryAdiabaticElimination} we will utilize a theory
for the adiabatic elimination of the unwanted (off-resonant) levels as shown
in Fig.~\ref{fig:FourLevels}. Thus using the basis of
Eq.~(\ref{eq:fourmodebasis}) we will develop an effective two-level
Hamiltonian which takes the form
\begin{equation}\label{eq:Heff-iSWAP-slow}
\eff{H}=\begin{bmatrix}
0 & \eff{g}\\
\eff{g} & \eff{\Delta}
\end{bmatrix} \ ,
\end{equation}
where ($g$$\ll$ $\Delta$) is required (see
appendix~\ref{app:TheIswapGate1}). The effective coupling is
found to be 
\begin{align}\label{eq:geff-ISWAP-slow}
\eff{g}\approx-\frac{\Gee{ab}{1}\Gee{bc}{2}\Gee{cd}{3}\Gee{da}{4}}{\Delta_1\Delta_2\Delta_3}\ ,
\end{align}
and the effective detuning of the two-level system is 
\begin{eqnarray} \label{eq:Delta-ISWAP-slow}
\Delta_{\rm{eff}}\approx \Delta_4+\frac{(\Gee{ab}{1})^2}{\Delta_1}-\frac{(\Gee{da}{4})^2}{\Delta_3} \ .
\label{eq:DeltaEffIn2LevelIniSWAP}
\end{eqnarray}
The time evolution of this system is given by the following equations for
the logical states,
\begin{align}\label{eq:iSWAP-slow-2Eqns}
\nonumber
\ket{a, 10}&\mapsto\cos(\eff{g}t)\ket{a, 10}-\rm{i}\sin(\eff{g}t)\ket{a, 01}\\
\ket{a, 01}&\mapsto\cos(\eff{g}t)\ket{a, 01}-\rm{i}\sin(\eff{g}t)\ket{a, 10}\ .
\end{align}

\begin{figure}\centering

  \includegraphics[width=0.5 \textwidth]{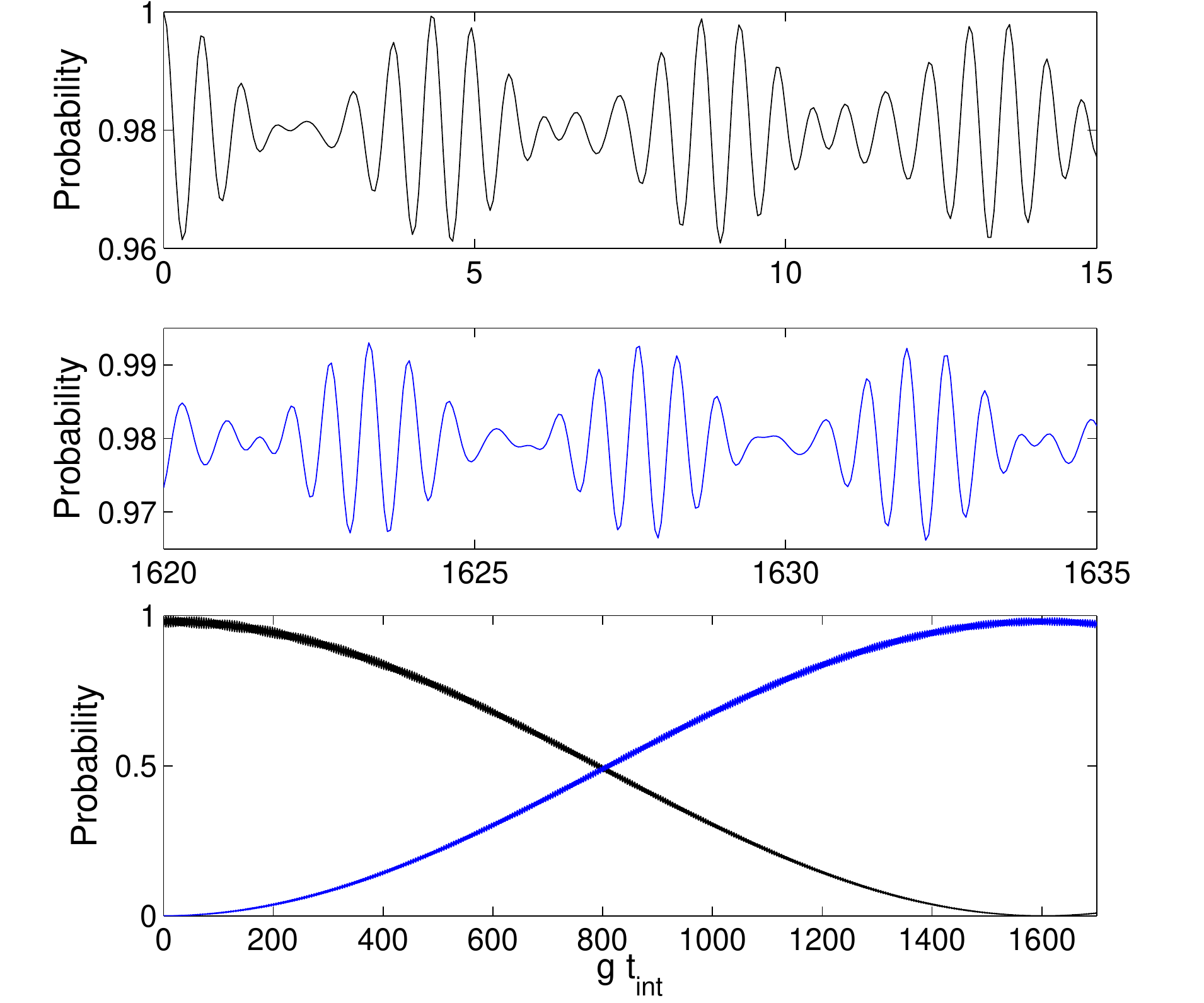}
                                                                        
  \caption{Population swapping in the model (10001). All
    couplings $g_j$ ($j=1,2,3,4$) are set to $g$, and the detunings
    $\Delta_i$ with $i=1,2,3$ are set to $\Delta$. The value of $\Delta_4$ is
    determined by Eq.~(\ref{eq:DeltaEffIn2LevelIniSWAP}).}\label{fig:iSWAPslow}
\end{figure}

This two-level system undergoes swapping of the states $|a1010\rangle$ and
$|a0101\rangle$ when the resonance condition is achieved by setting
$\Delta_{\rm{eff}}$ to zero in equation (\ref{eq:DeltaEffIn2LevelIniSWAP}).
This is illustrated in Fig.~\ref{fig:iSWAPslow} where excellent agreement
between an exact numerical calculation and the analytic treatment of
equations~(\ref{eq:iSWAP-slow-2Eqns}) is presented. Because of the finite
value of $\Delta$ chosen in Fig.~\ref{fig:iSWAPslow} a fine high frequency
oscillation can be seen. For larger detuning this oscillation becomes
smaller as the two-level approximation is realised more accurately.

\begin{figure}

  \includegraphics[width=0.45\textwidth]{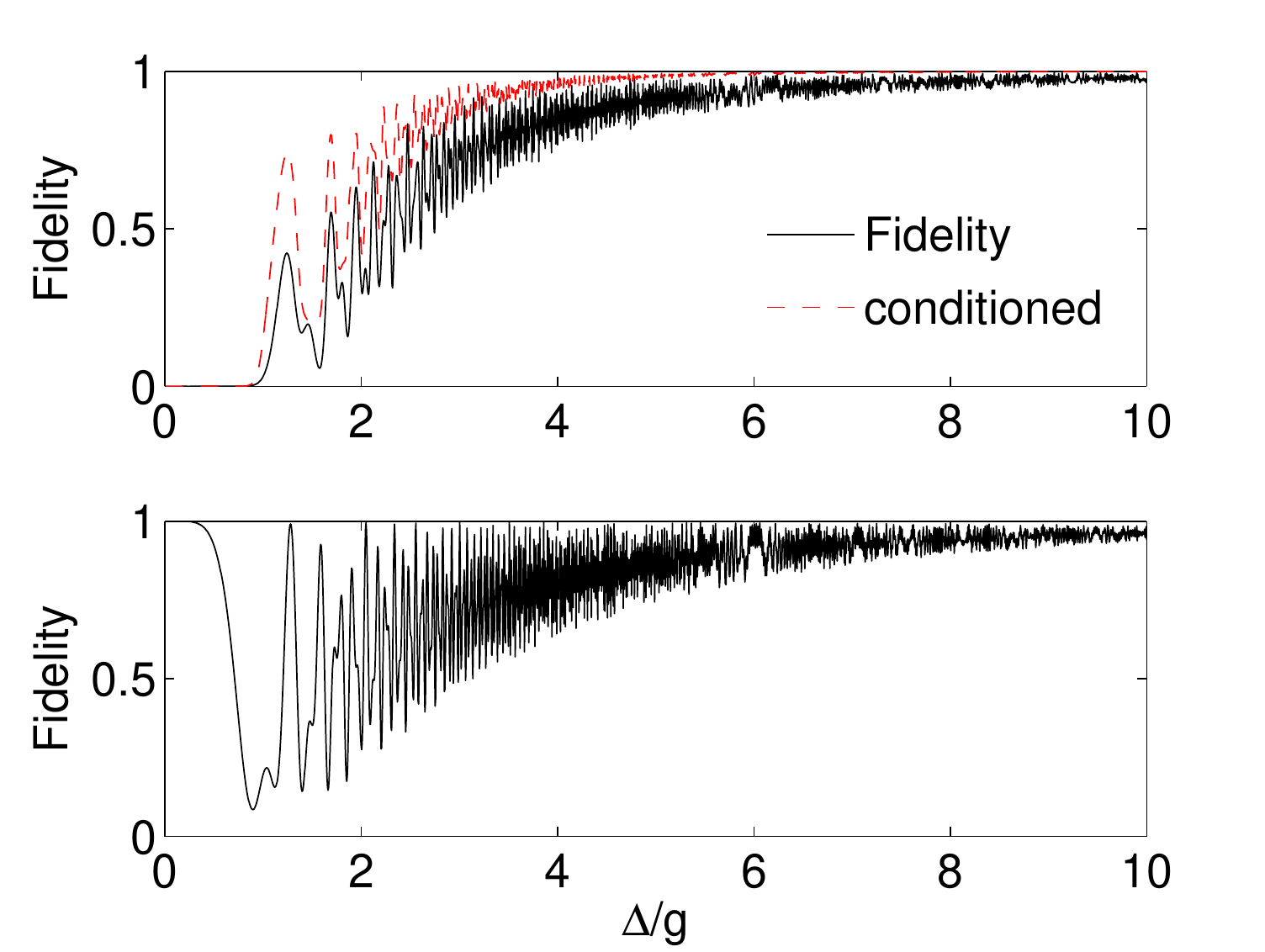}\llap{
 \parbox[b]{.005in}{(b)\\\rule{0ex}{.6in}}}\llap{
 \parbox[b]{.005in}{(a)\\\rule{0ex}{1.8in}}}
  
  \caption{(a) The fidelity (black line) of swapping the logical states $|a\ 10\rangle$ and $|a\
    01\rangle$ in the model (10001). Red-dashed line shows the conditional
    fidelity. (b) The fidelity of keeping the system initially in the logical state $|a\
    11\rangle$. Parameters of the detunings $\Delta$ and the coupling
    strengths $g$ are defined in Fig.~\ref{fig:iSWAPslow}.} \label{fig:M(10001)_fidelities} 
\end{figure}

The effective Hamiltonian~(\ref{eq:Heff-iSWAP-slow}) only connects the
states $|a\ 1010\rangle$ and $|a\ 0101\rangle$; the state $|a\ 1001\rangle$
is effectively governed instead by the full Hamiltonian~(\ref{eq:a1001
  Hamiltonian}), and the other logical state of the system, $\ket{00}$, is
unchanged. Considering the parameters in Fig.~\ref{fig:iSWAPslow} and at the
special time $|\eff{g}t|=\pi/2$, we obtain the following outputs for the
basic inputs (see Fig.~\ref{fig:M(10001)_fidelities}):
\begin{equation}
\begin{tabular}{ccc}
Input && Output\\
\hline
$\ket{00}$ && $\ket{00}$\\
$\ket{01}$ && $\rm{i}\ket{10}$\\
$\ket{10}$ && $\rm{i}\ket{01}$\\
$\ket{11}$ && $\ket{11}$
\end{tabular}\label{eq:iSWAP-slow-logic-table}
\end{equation}
This realises an i\textsc{swap} gate and we can represent the above mappings as
the table
\begin{equation}
  \mbox{i\textsc{swap}} \equiv \left[ \begin{array}{rrrr}%
1& & & \\
 &0&\rm{i}& \\
 &\rm{i}&0& \\
 & & &1\\
\end{array}\right] \ .
\end{equation}
The i\textsc{swap} gate is a universal gate when combined with single qubit
rotations. For example it can be directly related to the \textsc{cnot} gate (also a
universal gate) by means of a quantum circuit as shown in \cite{Schuch}.
Recently, it is shown that the i\textsc{swap} gate can be very useful for applications 
in quantum information process QIP and quantum computing. For example, it is
reported that the replacement of the standard \textsc{cnot} gate by the i\textsc{swap} gate
provides more efficient, simpler, and faster way of generating cluster states 
\cite{Tanamoto}, which  play a crucial role in the so-called one-way quantum
computation approach \cite{Raussendorf2, Raussendorf3}.

%
%

\subsection{A three-level behaviour}
\label{sec:3LevelBehavior}

Model (11001) has an intermediate resonant state. The system takes the form of
a double $\Lambda$ system with a common initial and final state. Since
$|b\ 0010\rangle$ is taken to be resonant, we apply the adiabatic
elimination theory (as shown in appendix~\ref{app:TheIswapGate2}) to obtain
an effective three-level system in the reduced space of states
$\{|a\ 1010\rangle,|b\ 0010\rangle,|a\ 0101\rangle\}$.  The effective
couplings are found to be
\begin{eqnarray}
g_{\rm{eff}}^{(1)}=g_1^{\rm{ab}} ,\qquad
g_{\rm{eff}}^{(2)}
\approx\frac{g^{\rm{bc}}_2g^{\rm{cd}}_3g^{\rm{da}}_4}{\Delta_2\Delta_3} \ 
, \label{eq:effective coupling in 11001} 
\end{eqnarray}
and the effective detunings
\begin{eqnarray}
\nonumber
\Delta_{\rm{eff}}^{(1)}&\approx&\Delta_1-\frac{(\Gee{bc}{2})^2}{\Delta_2}\ ,\\
\Delta_{\rm{eff}}^{(2)}&\approx&\Delta_4-\frac{(\Gee{da}{4})^2}{\Delta_3}\
.\label{eq:resonance conditions in 11001} 
\end{eqnarray}
The resulting Hamiltonian takes the form 
\begin{equation}
H_{\rm{eff}}=\left[\begin{array}{ccc}0&g_{\rm{eff}}^{(1)}&0\\
                                      g_{\rm{eff}}^{(1)}&\Delta_{_{\rm{eff}}}^{(1)}&g_{\rm{eff}}^{(2)}\\
                                    0&g_{\rm{eff}}^{(2)}&\Delta_{_{\rm{eff}}}^{(2)}  
                 \end{array}\right] \ .
\label{eq:Three-level eff_Ham}
\end{equation}

\begin{figure}\centering

  \includegraphics[width=.4\textwidth]{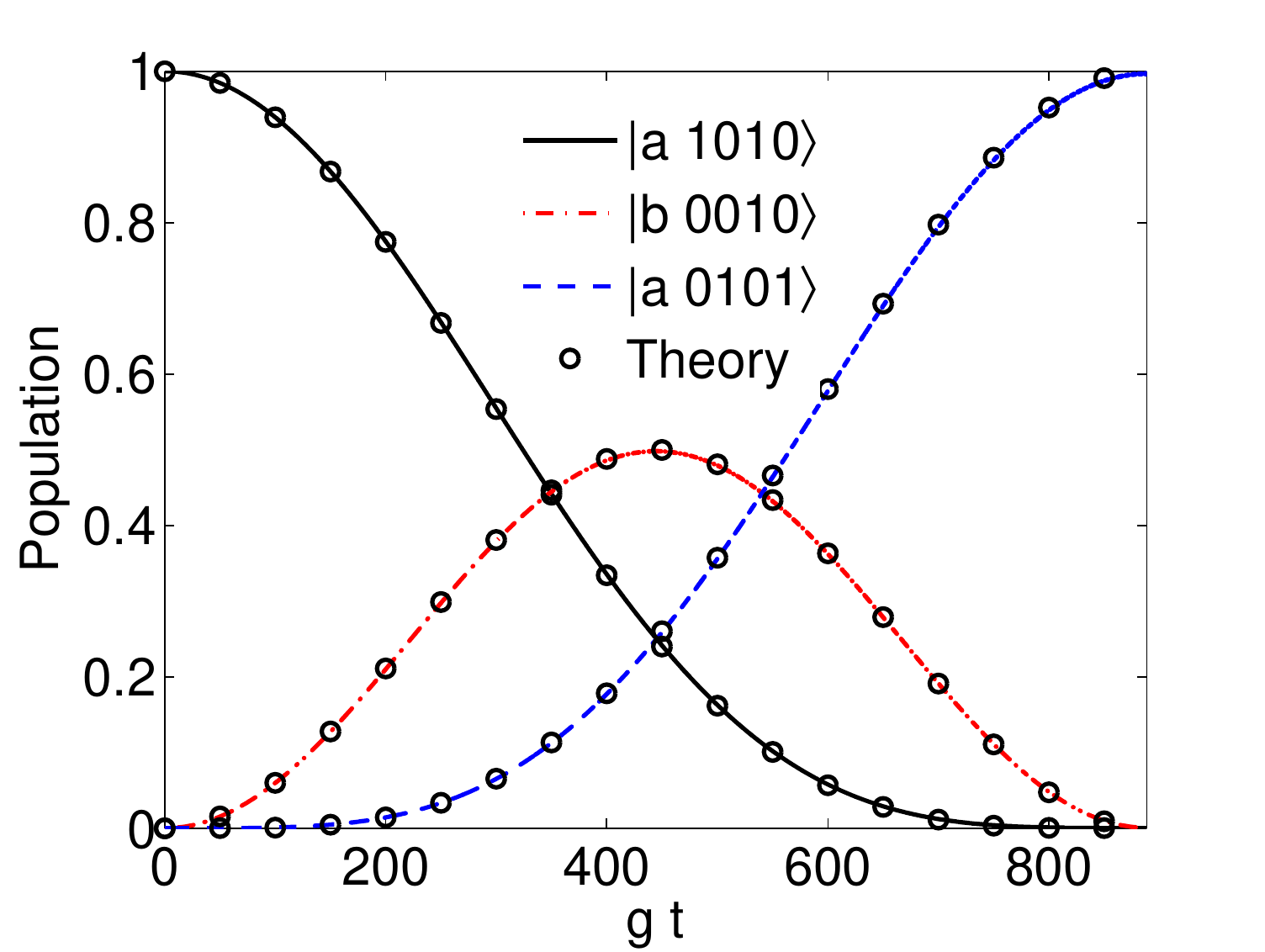}\llap{
 \parbox[b]{6in}{(a)\\\rule{0ex}{1.1in}}}

\includegraphics[width=.4\textwidth]{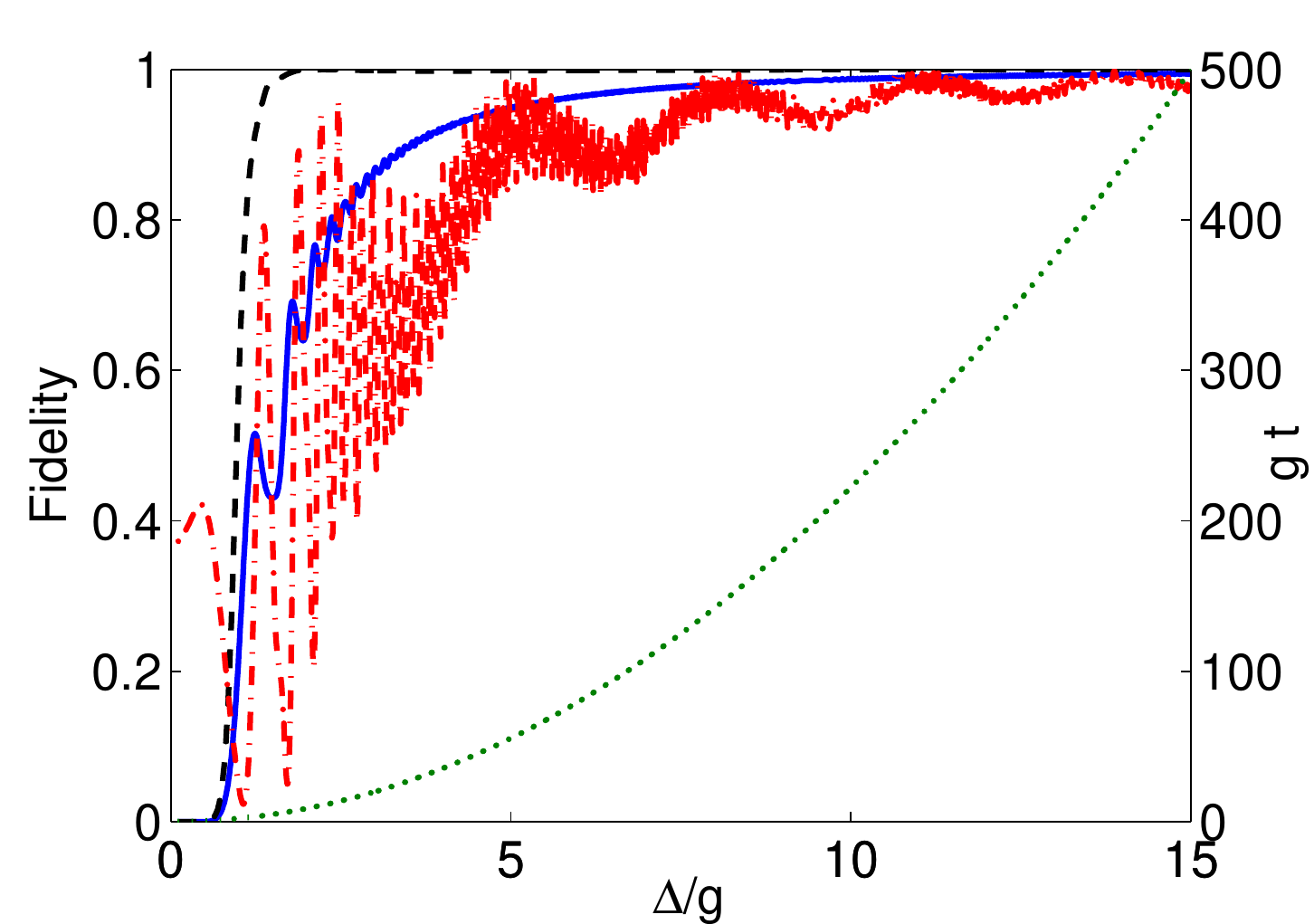}\llap{
 \parbox[b]{6in}{(b)\\\rule{0ex}{1in}}}
  
  \caption{[Color online] A three-level behaviour with the states $|c\rangle$
    and $|d\ \rangle$ chosen to be highly-detuned (see 
    Fig.~\ref{fig:FourLevels}). Parameters: the coupling constants $g_{2,3,4}$
    are all set to $g$, and the detunings $\Delta_{2,3}$ are set to
    $\Delta$. (a) the probability of the states $|a\ 1010\rangle$, $|b\
    0010\rangle$, and $|a\ 0101\rangle$ with $\Delta=20 g$. (b) the fidelity of
    mapping $|a\ 10\rangle$ to $|a\ 01\rangle$ (blue-solid line), and of
    keeping $|a\ 11\rangle$ in its initial state(red dotted-dashed line). The
    black-dashed and green-dotted lines represent the conditioned fidelity and
    the interaction time $gt_{\rm{int}}$,
    respectively.} \label{fig:M11001_swapping_fidelities}
\end{figure}

The time evolution of this gate is shown in
Fig.~\ref{fig:M11001_swapping_fidelities}(a) and it is immediately apparent
that the population swapping happens much faster than the evolution seen in
Fig.~\ref{fig:iSWAPslow} for Model (10001). The time evolution of the
swapping follows the equations (for the initial state $|a,1010\rangle$)
\begin{eqnarray}
\nonumber|a,10\rangle\longrightarrow && 
[\frac{(\Gee{(1)}{\rm{eff}})^2}{\overline{g}^2}+\frac{(\Gee{(2)}{\rm{eff}})^2}{\overline{g}^2}\ 
\cos(\overline{g}t)]\ |a,10\rangle\\
&-&\rm{i}\frac{\Gee{(1)}{\rm{eff}}}{\overline{g}}\ \sin(\overline{g}t)\
|\Phi\rangle \label{eq:three_level_analytical_sol}\\
\nonumber&+&\frac{\Gee{(1)}{\rm{eff}} \Gee{(2)}{\rm{eff}}}{\overline{g}^2} \
[\cos(\overline{g}t)-1]\ e^{i\eta 
  t}\ |a,01\rangle\ ,
\end{eqnarray}
where
$\overline{g}=\sqrt{(\Gee{(1)}{\rm{eff}})^{2}+(\Gee{(2)}{\rm{eff}})^{2}}$ and
$|\Phi\rangle\equiv|\rm{b}\ 0010\rangle$. 

The spin-$J$ model predicts the proper values for the effective coupling
constants so that a complete qubit swapping can be achieved. That is, in the
previous three-level system, the general scaling expression \cite{Cook}
\begin{eqnarray}
\Gee{(n)}{\rm{eff}}=g_0\sqrt{n(N-n)} \label{eq:SequenceOfRabiFreq.}
\end{eqnarray} 
(where $g_0$ is a constant and in our case $N=3$ and $n=1,2$) suggests that
$|\Gee{(1)}{\rm{eff}}t|=|\Gee{(2)}{\rm{eff}}t|=\pi/\sqrt{2}$. Then, the
transformation $|a\ 1010\rangle \rightarrow$ $\exp(\rm{i}\eta t)|a\
0101\rangle$ can take place by setting $\overline{g}t=\pi$.  
%
A global phase $\eta t$ in the previous two- and three-level systems can be 
produced by following different proposals. Some examples will be suggested
when we discuss the single-qubit gate in Sec.~\ref{SingleQubitGates}. 

By substituting the parameters of the coupling constants and detunings in
the model (11001) into the Hamiltonian~(\ref{eq:a1001 Hamiltonian}) which
describes the time evolution of the initial state $|a\ 1001\rangle$, it is
noticeable that this state can be sufficiently forced to stay in its
initial state, as demonstrated by the red-dashed line in
Fig.~\ref{fig:M11001_swapping_fidelities}(b).

\subsection{Enhancement conditional on measurements}

In the scheme we have developed so far the atom plays the role of an ancilla
which simply ``enables'' the shuffling of energy between cavity modes.
However, if the proposed gate is slightly imperfect there will be a small
admixture of other atomic states. The role of a conditional measurement can
be to improve the quality of the final state (i.e.\ to improve the
fidelity). For example, let us suppose the state of the system is
\begin{align}\label{eq:cond-example-all}
|\Psi(t)\rangle&=c_1  |a\ 1010\rangle+ c'|c\ 0110\rangle
+c_2 |a\ 0101\rangle  + ...
\end{align}
Then if the atom exits the system in state $\k a$ we must project the
state (\ref{eq:cond-example-all}) onto the atomic state $\k a$ to
obtain the conditioned result 
\begin{align}\label{eq:cond-example-measured}
|\Psi(t)\rangle^{\prime}&\rightarrow \frac{c_1}{\sqrt{|c_1|^2 + |c_2|^2}}|a\ 1010\rangle
             + \frac{c_2}{\sqrt{|c_1|^2 + |c_2|^2}} |a\ 0101\rangle. 
\end{align}
Because of the renormalisation that takes place this conditionally enhances a
desired result (such as $|a\ 0101\rangle$). 

To a limited extent this makes the scheme probabilistic. However, the
probability of success in such a measurement is expected to be high and the
intent is that this measurement process simply enhances the result and
cleans up the wave-function. Red-dashed line in
Fig.~\ref{fig:M(10001)_fidelities}(a) and black-dashed line in
Fig.~\ref{fig:M11001_swapping_fidelities}(b) show significant improvement of
fidelity in the models (10001) and (11001).


\subsection{Decoherence process}
\label{sec:M10101-decoherence}

The time evolution of the previous systems can be governed by Liouville's
equation
\begin{eqnarray}
\frac{\partial}{\partial t}\ \rho=-\rm{i} \left[H,\ \rho\right]+\mathcal{L}\
\rho\ ,\label{eq:MasterEquation}
\end{eqnarray}
where $\rho$ is the density operator of the atom-field system 
and the so-called Liouvillian operator $\mathcal{L} \rho$ describes the
dissipative mechanisms in the system. The general Lindblad form of the
Liouvillian operator $\mathcal{L} \rho$ can be expressed as
\cite{Lindblad}
\begin{eqnarray}
\mathcal{L}\ \rho&=&\sum \limits_{i} \frac{\eta_{(i)}}{2}\
([L_{(i)}\ \rho\ ,\ L^{\dag}_{(i)}]+[L_{(i)}\ ,\ \rho \ L^{\dag}_{(i)}])\ ,
\label{eq:Liouville operator}
\end{eqnarray} 
where $\eta$ represents the loss of population which can be due to either the 
spontaneous emission $\Gamma$ or to the cavity field rate $\kappa$. The
operators $L$ and $L^{\dag}$ are the corresponding system operators. More
explicitly, in the presence of the atomic decay $L$ and $L^{\dag}$ can be
replaced by the atomic operators $\sigma_{-}$ and $\sigma_{+}$, and in the
case of the cavity decay they are represented by the field operators $a$ and
$a^{\dag}$. 

Given the initial states to be either $|a\ 1010\rangle$ or $|a\
0101\rangle$, we can also investigate the influence of the atomic and
photonic relaxations by considering the following Hamiltonian
\begin{eqnarray}
H^{\prime}=H-\rm{i}\frac{\kappa}{2}\sum \limits_{i=1}^{4}\
a^{\dag}_{i}a_{i}\ -\rm{i}\frac{\Gamma}{2} (|b\rangle \langle b|+|d\rangle 
\langle d|)\ ,
\label{eq:ConditionalHamiltonian} 
\end{eqnarray}
where $H$ is the original Hamiltonian of the system in the absence of any
decay. 

\begin{figure}

\subfigure[]{\includegraphics[width=0.5 \textwidth]{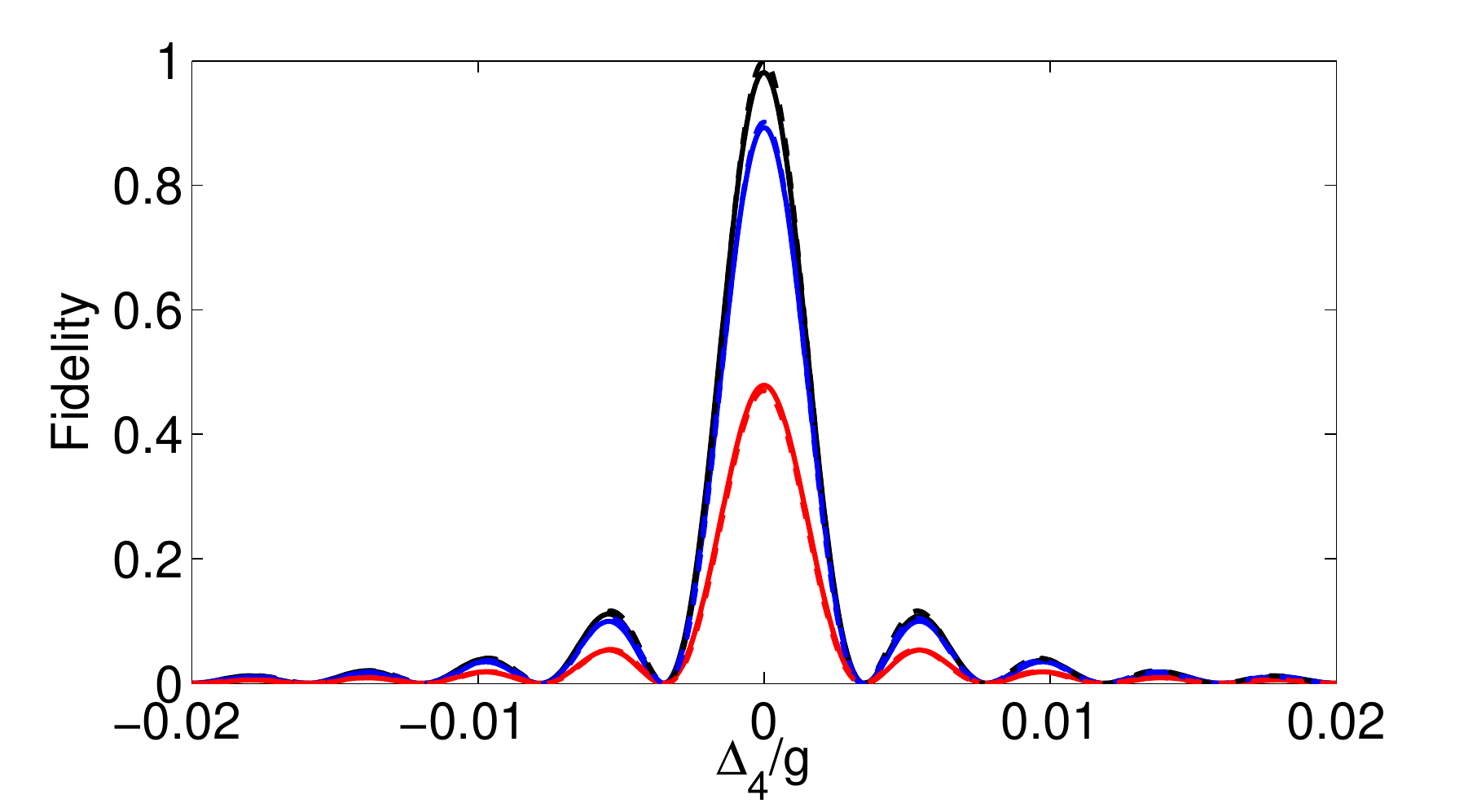}}
\subfigure[]{\includegraphics[width=0.5 \textwidth]{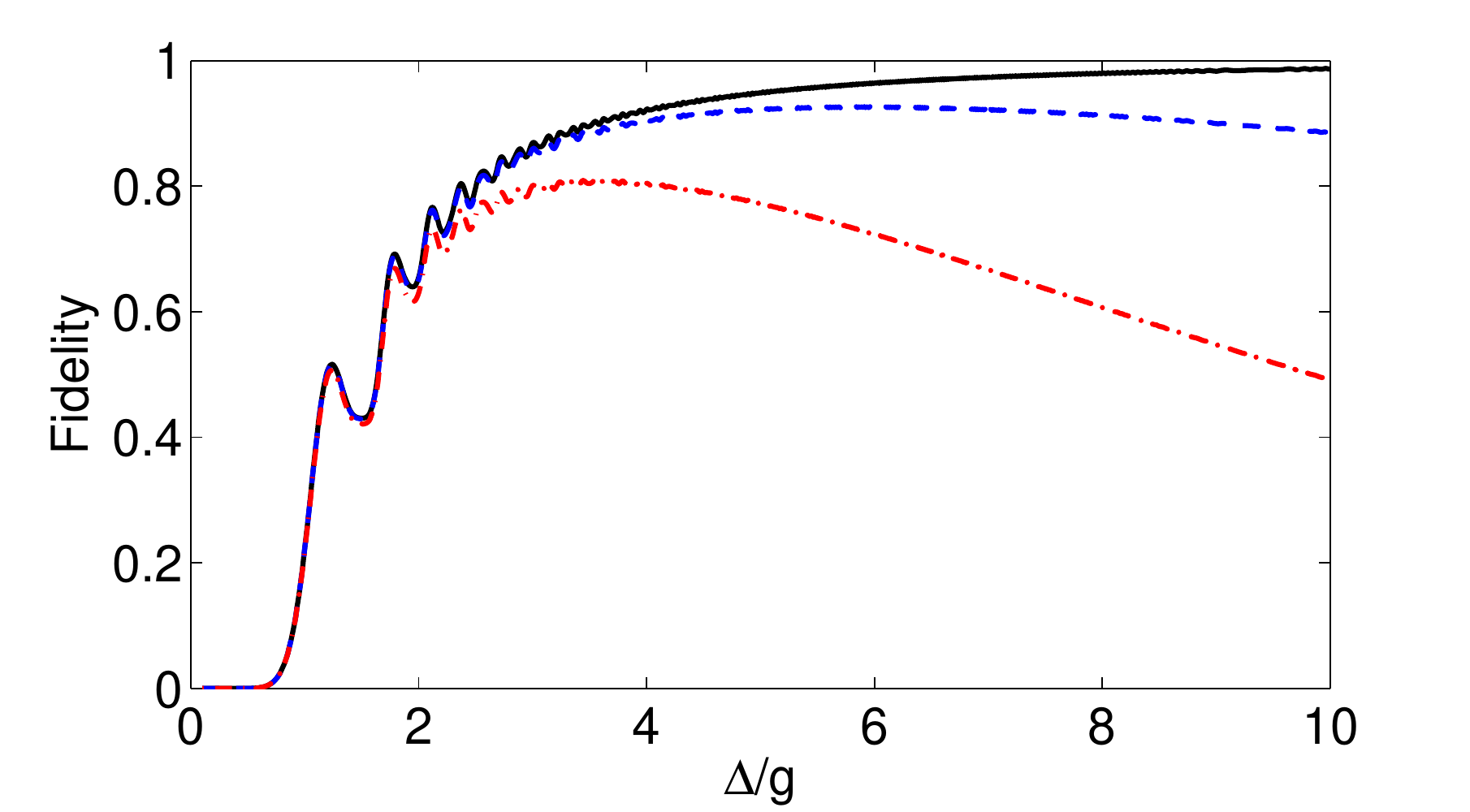}}
\caption{The influence of the dissipative mechanisms on the qubit state $|a\
  10\rangle$ or $|a\ 01\rangle$ in the models (10001) and (11001). (a) in the
  model (10001), the three curves corresponding to the values of photonic
  decays $\kappa \sim (0, 0.033, 0.24)g_{\rm{eff}}$ with $\Delta=10 g$. The
  solutions by theory (dashed lines) and simulation (solid lines) are highly
  matched. (b) in the model (11001) The three curves corresponding to the
  values of the atomic decay rates $\Gamma \sim (0, 0.1, 0.75)
  \bar{g}/\sqrt{2}$ with $\kappa=0$, or to the values of photonic decays
  $\kappa \sim (0, 0.025, 0.185)\bar{g}/\sqrt{2}$ when no atomic decay rate is 
  considered. The three curves in all these plots represent (100, 90, 50)\%
  fidelity.}\label{fig:CavityDecayIn10001}
\end{figure}

This procedure is valid under the condition that no photon is
detected \cite{Dalibard, Molmer, Zoller}, and the Shore's method \cite{Shore}
can be reapplied to produce the damped N-level configurations. We have seen
previously that the models (10001) and (11001) are capable to realise the
i\textsc{swap} gate. Starting with the model (10001), the time evolution of the state
$|a\ 0101\rangle$ in the strong coupling regime is 
\begin{eqnarray}
c_{a01}(t)&=&-\rm{i}\frac{g_{\rm{eff}}}{\tilde{g}} \ \ e^{-\kappa t}\
\sin(\tilde{g}t)\ e^{-\rm{i}\Delta_{\rm{eff}}t/2}\ ,\label{eq:iSWAP (10001) 
  exact sol}
\end{eqnarray}
where $\tilde{g}=\sqrt{g_{\rm{eff}}^2+(\Delta_{\rm{eff}}/2)^2}$. The norm of 
the system shows that it decays with the rate
$(2\kappa)$. Figure~\ref{fig:CavityDecayIn10001}(a) demonstrates
the fidelity for different values of $\kappa$. As we have mentioned before, 
the i\textsc{swap} gate formed from the model (10001) is very slow gate and, therefore, it is
very sensitive to photonic decay rates.

In the case of the model (11001) and taking into account the atomic and
cavity decay rates in the effective Hamiltonian~(\ref{eq:Three-level eff_Ham}),
the eigenvalues for this effective Hamiltonian can be determined, under the
condition $4\bar{g}\geq (\kappa-2\Gamma)$ and with vanishing effective
detunings, as  
\begin{eqnarray}
\lambda_1=-\kappa\ ,\ \ 
\lambda_{2,3}=-(\frac{3\kappa+2\Gamma}{4})\pm \rm{i}\ {\lambda}\ ,
\end{eqnarray}
where $\lambda=\{ \bar{g}^2 -\frac{1}{4}(\kappa-2\Gamma)^2 \}^{1/2}$, 
$\bar{g}=\sqrt{(\Gee{(1)}{\rm{eff}})^2+(\Gee{(2)}{\rm{eff}})^2}$, and
$\Gee{(1)}{\rm{eff}}$ and $\Gee{(2)}{\rm{eff}}$ are the effective coupling
constants. The corresponding eigenvectors can be found and then the time
evolution of the logical state $|a\ 01\rangle$ can be expressed as 
\begin{widetext}
\begin{eqnarray}
c_{a01}(t)=\frac{\Gee{(1)}{\rm{eff}}\Gee{(2)}{\rm{eff}}}{{\bar{g}}^2}  \
\exp(-\kappa\ t)\
\left\{-1+\left[\cos({\lambda}t)-(\frac{\kappa-2\Gamma}{4{\lambda}}) \
  \sin({\lambda}t) \right] \ \exp(\frac{\kappa-2\Gamma}{4}\ t) \right\}\
,\label{eq:FastIswapPopulationLoss} 
\end{eqnarray}
for the initial condition that the system is completely set in the initial state
$|a\ 1010\rangle$ at $t=0$, i.e.\ $c_{a10}(0)=1$.
\end{widetext}

In Fig.~\ref{fig:CavityDecayIn10001}(b) we consider the model (11001) and
measure the fidelity at different values of $\kappa$.
Figure~\ref{fig:CavityDecayIn10001} shows that the impact of cavity
field relaxation is less in the model (11001) when compared to the model
(10001), and this is because of the improvement in the qubit states speed.

In the case of the initial state to be $|a\ 0110\rangle$, this state decays
due to the cavity relaxation and the population loss follows
$|c_{a00}|^2=\exp(-2\kappa t_{\rm{int}})$, where the interaction time
$t_{\rm{int}}$ is same to the models (10001) and (11001). In
the final state $|a\ 1001\rangle$ we directly use the master
equation~(\ref{eq:MasterEquation}) to investigate the effect of atomic and
photonic dampings in either the model (10001) or the model (11001).

\begin{figure}
  \centering
  \subfigure[]{\includegraphics[width=\bmgstdfigwidth]{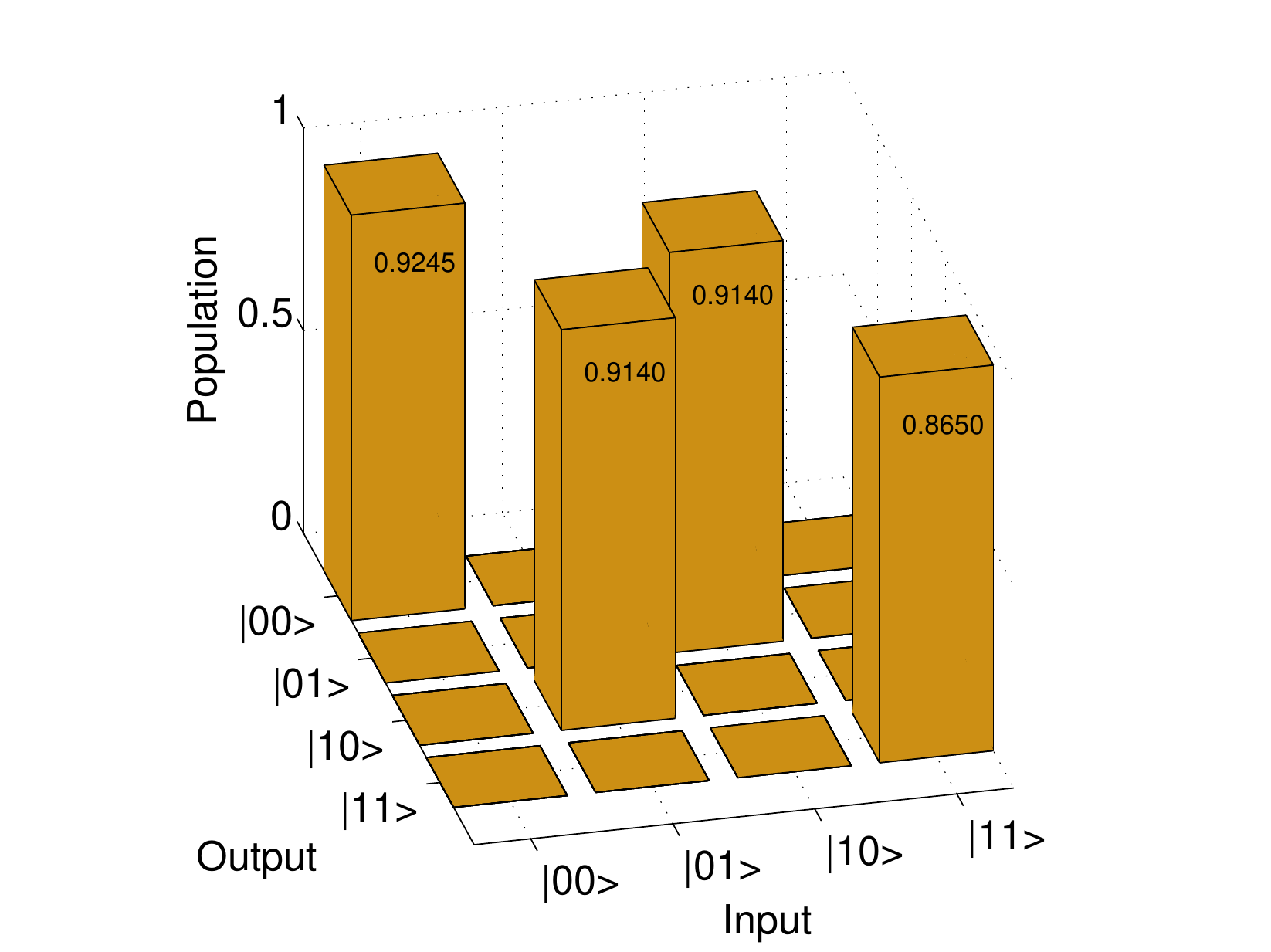}}
  \subfigure[]{\includegraphics[width=\bmgstdfigwidth]{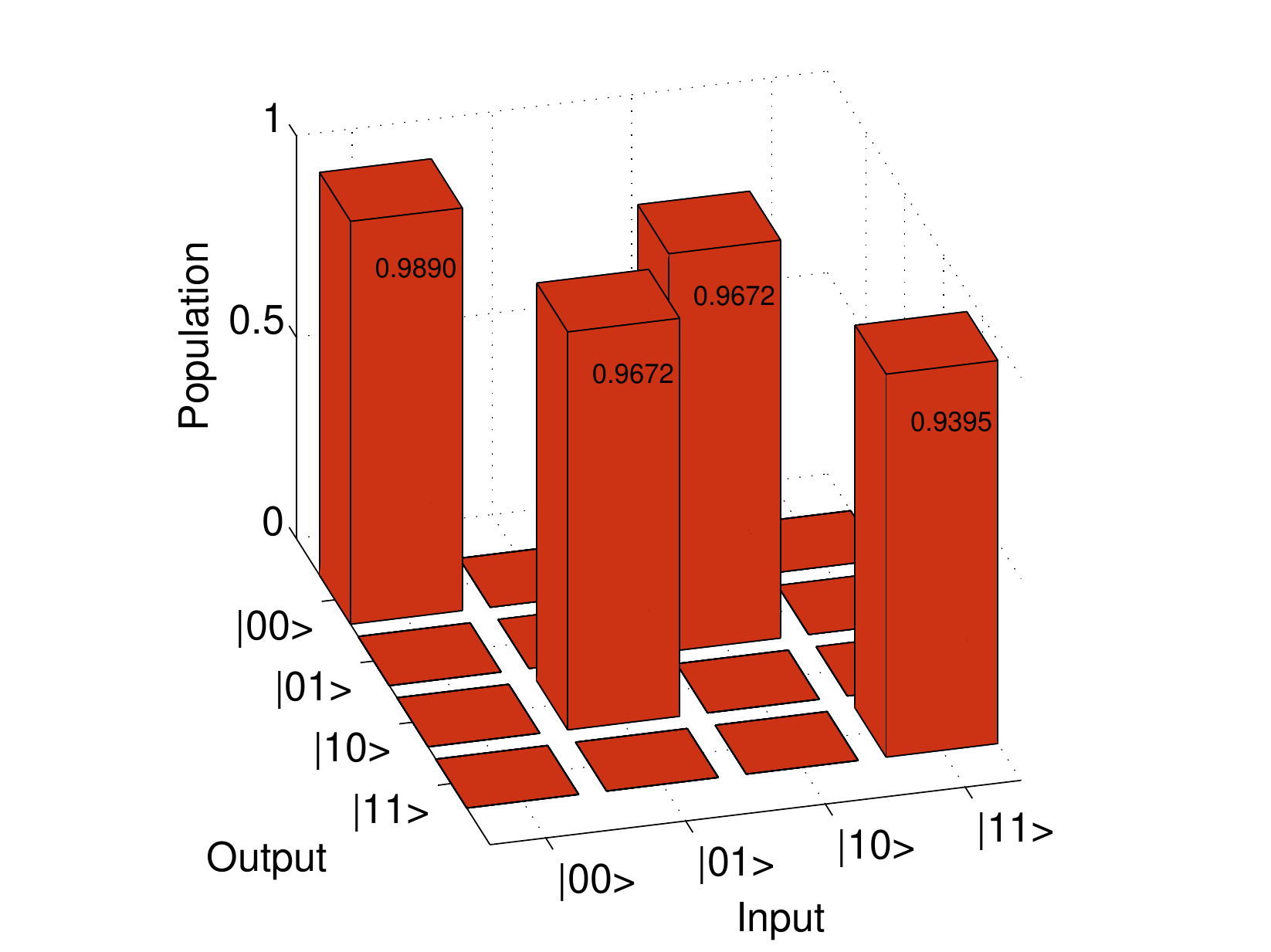}}
  \caption{Truth table of the numerically simulated i\textsc{swap} gate with
    $\Delta=10\ g$ in the presence of decoherence processes. (a)
  in the configuration (10001). (b) in the model (11001). Parameters: the
  coupling constant is approximate $g/2\pi=50$ kHz, $\Gamma/g \sim 10^{-4}$,
  and $\kappa/g \sim
  2.5\times10^{-5}$.} \label{fig:iSWAP_GatesWithExper.Parameters} 
\end{figure}

We are now in the position to subject all qubit states in the i\textsc{swap} 
gate to experimental values including the coupling strength $g$, the atomic
decay $\Gamma$, and the photonic decay rate $\kappa$ so that the performance
of the two-qubit dual-rail CQED gate can be practically tested. Considering the
microwave cavity-QED experiment in \cite{Haroche5}, highly excited
Rydberg atoms (typically $^{85}$Rb) with a radiative time $\tau_{rad}\sim 30$
ms have been used to interact with a superconducting cavity with $Q\mapsto
4\times 10^{10}$. The photon lifetime inside the cavity is in order
$\tau_{ph}\sim 130$ ms, and the coupling strength is around $g/2\pi\sim 50$
kHz. By setting $\Delta=10$ g, this corresponds to
cavity-atom interaction time $t_{int}\mapsto 5$ ms in the configuration
(10001), and $t_{int} \mapsto \frac{1}{\sqrt{2}}$ ms in the configurations
(11001). The quantity $\tau_{ph}/t_{int}$ shows that the last configuration
is much better for QIP applications with the present cavity QED techniques. 
Plots in Fig.~\ref{fig:iSWAP_GatesWithExper.Parameters} show the population
loss in the i\textsc{swap} gate realised by the configurations (10001) and (11001) with
considering the above values of the parameters $g$, $\Gamma$, and $\kappa$.


\section{Three qubit gates and rotations} 
\label{sec:ThreeAndOneQubitGates}

So far we have described a two-qubit entangling gate. However, to have
a universal set of gate operations it is necessary to have single
qubit rotations (or single qubit gates). We also present below a three
qubit gate, where full details will be presented
in appendices~\ref{sec:FastFredkinGate} and~\ref{NotGate}. 

\subsection{Three qubit gate}

The three qubit scheme of Fig.~\ref{fig:FredkinLevels} contains
dual-mode qubits with a repeated mode at $\omega_1$. The partner mode
for $\omega_1$ is not shown as it is unaffected by the logic process.
The remaining pairs are $\omega_{2,3}$ and $\omega_{5,6}$, (a mode
with the label $\omega_4$ is omitted in the following analysis to
avoid confusion). The consequence of the repeated mode~1 is that the
chain only completes if mode~1 is present (in which case the
excitation is absorbed and then re-emitted) and if modes~3 and 5 have
excitations present (which implies necessary empty modes~2 and~6 in the
dual-mode qubit basis). The result of all this is that the qubit~1 acts as
a control qubit which swaps the qubits present in logical qubits 2 and
3: i.e.\ we have the logic needed for a Fredkin gate \cite{Fredkin}. In a
Fredkin gate we aim for logical qubits to be mapped as follows:

\begin{equation}
\begin{tabular}{ccc}
Input && Output\\
\hline
$\ket{000}$ && $\ket{000}$\\
$\ket{001}$ && $\ket{001}$\\
$\ket{010}$ && $\ket{010}$\\
$\ket{011}$ && $\ket{011}$\\
$\ket{100}$ && $\ket{100}$\\
$\ket{110}$ && $\ket{101}$\\
$\ket{101}$ && $\ket{110}$\\
$\ket{111}$ && $\ket{111}$\\
\end{tabular}
\end{equation}


The Fredkin gate is a universal gate owning important
properties which set up the general principles of logic gates and circuits in
both classical and quantum computing \cite{Fredkin}. This gate swaps the
second and the third qubits if the first qubit is $|1\rangle$, otherwise, all
qubits remain unchanged. Two examples of quantum circuits generating this gate
can be considered. Firstly, it is observable that the Fredkin gate is nothing but the
controlled \textsc{swap} gate \cite{Nielsen}. Secondly, since the \textsc{swap} gate is equivalent to three
\textsc{cnot} gates, we can see that the Fredkin gate is a combination of the Toffoli
gate together with two \textsc{cnot} gates \cite{Barnett3}.
\subsection{Two- and three-level configurations}

  \begin{figure}
    
 \subfigure[]{\includegraphics[width=\bmgstdfigwidth]{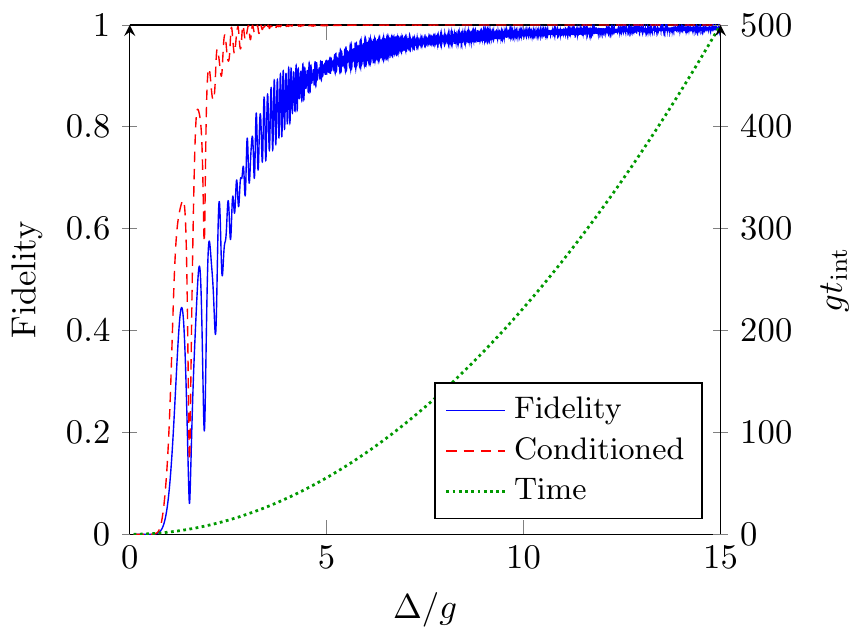}}
 \subfigure[]{\includegraphics[width=\bmgstdfigwidth]{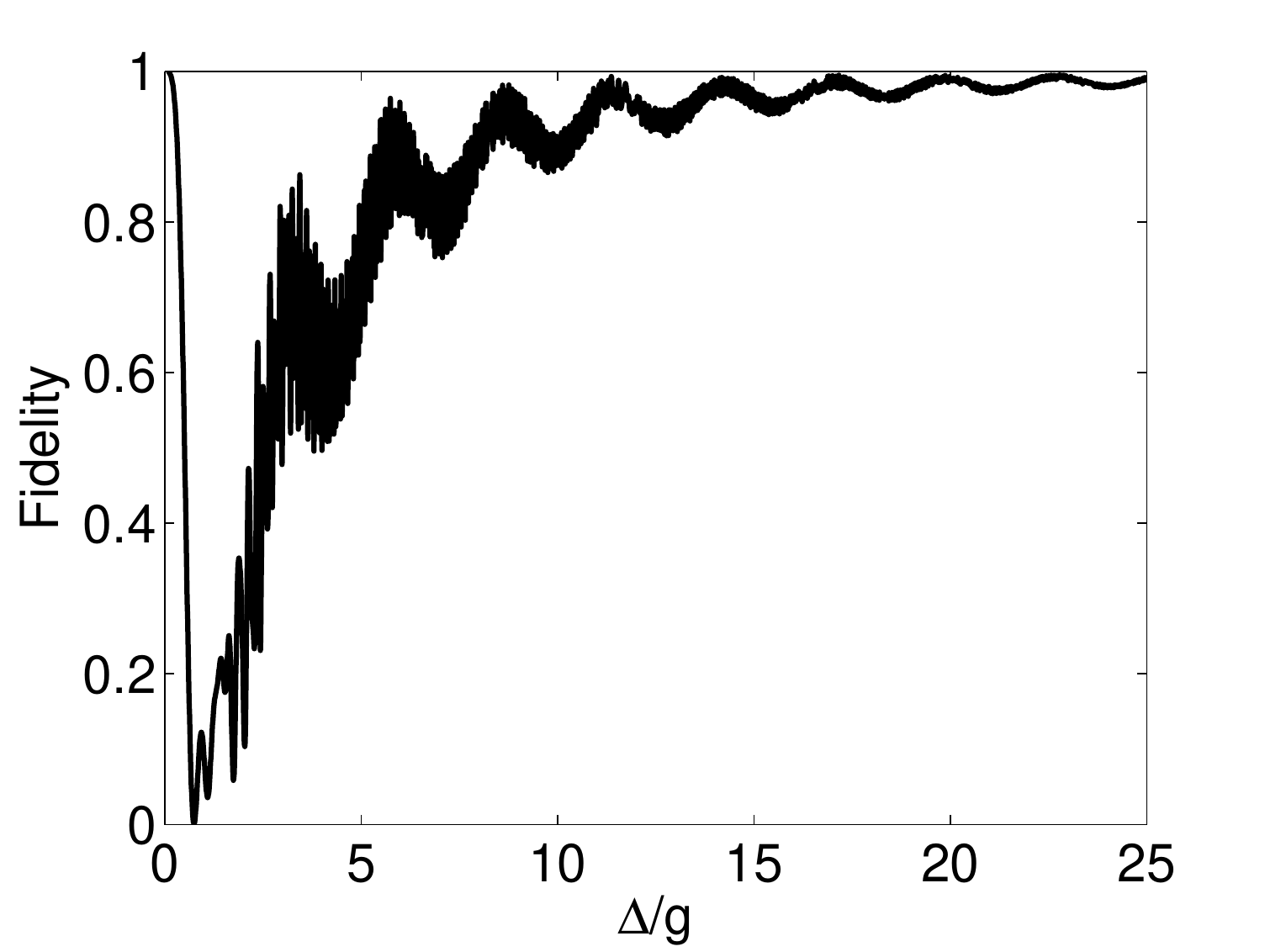}}
  \caption{Fidelity for the fast Fredkin gate shown as a function of the
    detuning $\Delta$ where we again take $\Delta_j \rightarrow \Delta $ for
    $j=1, 2, 4, 5$ with $\Delta_3=0$ and $\Delta_6\sim 0$ (given by
    equation~(\ref{eq:Fredkin-Delta-eff}) set to zero). 
    The initial state is (a) $|a\ 101\rangle$, (b) $|a\ 100\rangle$.
    For the couplings we
    take $ g_j \rightarrow g $. The dashed line in (a) shows the interaction time
    $\bar{g}t_\mathrm{int} =\pi$ as found from the three-state
    model~\ref{sec:3LevelBehavior}.} 

\label{fig:FastFredkinFidelity}
\end{figure}

  For a de-excited atom interacting with multimode
  cavities, the realisation of the Fredkin gate at the interaction time
  $gt_{\rm{int}}$ can be achieved when the transformation $|a\ 101\rangle
  \leftrightarrow |a\ 110\rangle$ is made and the remaining logical qubits
  are in their initial states. Once again, Shore's method plus the spin-$J$ 
  model provide useful tools to find out several configurations that are
  capable of swapping the states $|a\ 101\rangle$ and $|a\
  110\rangle$. However, to meet the truth table of the Fredkin gate, and
  since there is a presence of a repeated mode~1 in addition to certain over-shot
  states (see appendix~\ref{sec:FastFredkinGate}) it is noticed that the only
  possible configurations realising the gate are the models $(1000001)$ and
  $(1001001)$. In the model $(1000001)$ we set all states other than the
  states $|a\ 101\rangle$ and $|a\ 110\rangle$ to be far-resonance, and in the
  model $(1001001)$ a further atomic state $|d\rangle$ is allowed to be resonant. 
  The full details of the analysis of the former model (1000001) are not
  presented here, it can be found in \cite{Barry1} where we also show that an
  excellent speed-up is obtained by chosen an intermediate resonant
  energy level, i.e.\ the model (1001001). In this case the effective
  Hamiltonian reduces to a three-level system again, where (see
  appendix~\ref{sec:FastFredkinGate}) the effective coupling
  constants are 
\begin{align}\label{eq:Fredkin-g-eff}
\Gee{(1)}{\rm{eff}} &=
\frac{\Gee{ab}{1}\Gee{bc}{2}\Gee{cd}{3}}{\Delta_1\Delta_2},& \Gee{(2)}{\rm{eff}}
& = \frac{\Gee{de}{1}\Gee{ef}{5}\Gee{fa}{6}}{\Delta_4\Delta_5}\ ,
\end{align}
and the effective detunings are
\begin{equation}\label{eq:Fredkin-Delta-eff}
\begin{split}
&\Delta_1^{\mathrm{eff}}\approx\Delta_3 +
\frac{\left(\Gee{ab}{1}\right)^2}{\Delta_1} -
\frac{\left(\Gee{cd}{3}\right)^2}{\Delta_2}-\frac{\left(\Gee{de}{1}\right)^2}{\Delta_4}\,,\\
&\Delta_2^{\mathrm{eff}}\approx\Delta_6 -
\frac{\left(\Gee{af}{6}\right)^2}{\Delta_5}\ .
\end{split}
\end{equation}

Given the initial state to be either the logic $|a\ 101\rangle$ or $|a\
110\rangle$, equations of motion in~(\ref{eq:three_level_analytical_sol})
and the sequence of effective Rabi frequencies
in~(\ref{eq:SequenceOfRabiFreq.}) can be reused to swap the qubit states
$|a\ 10\rangle$ and $|a\ 01\rangle$. A high fidelity result can be obtained
for sufficient detunings, Fig.~\ref{fig:FastFredkinFidelity}(a) shows how this
increases, and also how the use of measurement of the ancilla atom strongly
improves the result for the quite low value of $\Delta/g\gtrsim 3$. The same
Fig.~shows the interaction time which is fairly high for large $\Delta$
($gt\sim 500$ for $\Delta=15g$) but reduces substantially near $\Delta=5g$.
The fidelity for keeping the qubit state $|a\ 100\rangle$ in its initial
state at the same interaction time above has been illustrated in
Fig.~\ref{fig:FastFredkinFidelity}(b). For other qubit states, large detunings
for all atomic levels except $|a\rangle$ and $|d\rangle$ ensure an efficient
confinement of the populations in desired levels so that the three-qubit
Fredkin gate is built.


\subsection{Decoherence in fast Fredkin gate}
Considering the parameters of the coupling constants and the detunings in the
model (1001001) given by
Eqs.~(\ref{eq:Fredkin-g-eff}, \ref{eq:Fredkin-Delta-eff}), we can use the
master equation~(\ref{eq:MasterEquation}) to address the influence 
of atomic and photonic relaxations on the Fredkin gate. Furthermore, we can
use the conditional Hamiltonian~(\ref{eq:ConditionalHamiltonian}) to find
analytic solutions for certain qubit states in Fredkin gate. For example,
the damping due to atomic and photonic decays can be studied as follows. 
Under the strong coupling regime, the eigenvalues of the damped three-level
behaviour (1001001) can be given as 
\begin{eqnarray}
\lambda_1=-3\kappa/2\ ,\ \ \lambda_{2,3}=-(\frac{5\kappa+2\Gamma}{4})\pm
\rm{i} {\lambda}\ ,
\end{eqnarray}
where $\lambda=\bigg(\bar{g}^2-\frac{1}{4}(\kappa-2\Gamma)^2\bigg)^{1/2}$, 
$\bar{g}=\sqrt{(\Gee{(1)}{\rm{eff}})^2+(\Gee{(2)}{\rm{eff}})^2}$, and
$\Gee{(1)}{\rm{eff}}$ and $\Gee{(2)}{\rm{eff}}$ are given by
Eq.~(\ref{eq:Fredkin-g-eff}) and the effective detunings all are set to
zero. The time evolution of the coefficient $c_{a110}$, with the initial
conditions $c_{a101}(t=0)=1$, $c_{d}(0)=0$, and $c_{a110}(0)=0$, reads
\begin{widetext}
\begin{eqnarray}
c_{a110}=\frac{\Gee{(1)}{\rm{eff}}\Gee{(2)}{\rm{eff}}}{{\bar{g}}^2}\
\exp(-\frac{3\kappa}{2}\ t)\ \{-1+\left[\cos({\lambda}t)-(\frac{\kappa-2\Gamma}{4{\lambda}})\
  \sin({\lambda}t)\right]\ \exp(\frac{\kappa-2\Gamma}{4}\ t)\} \ 
.\label{eq:FastFredkinPopulationLoss} 
\end{eqnarray}
Figure~\ref{fig:DampedFredkinGate} demonstrates the population loss by either
the spontaneous emission $\Gamma$ or the cavity field decay $\kappa$, and the
damped Fredkin gate under the recent QIP techniques. 
\begin{figure}
  \centerline{  
  \subfigure[]{\includegraphics[width=.35\textwidth]{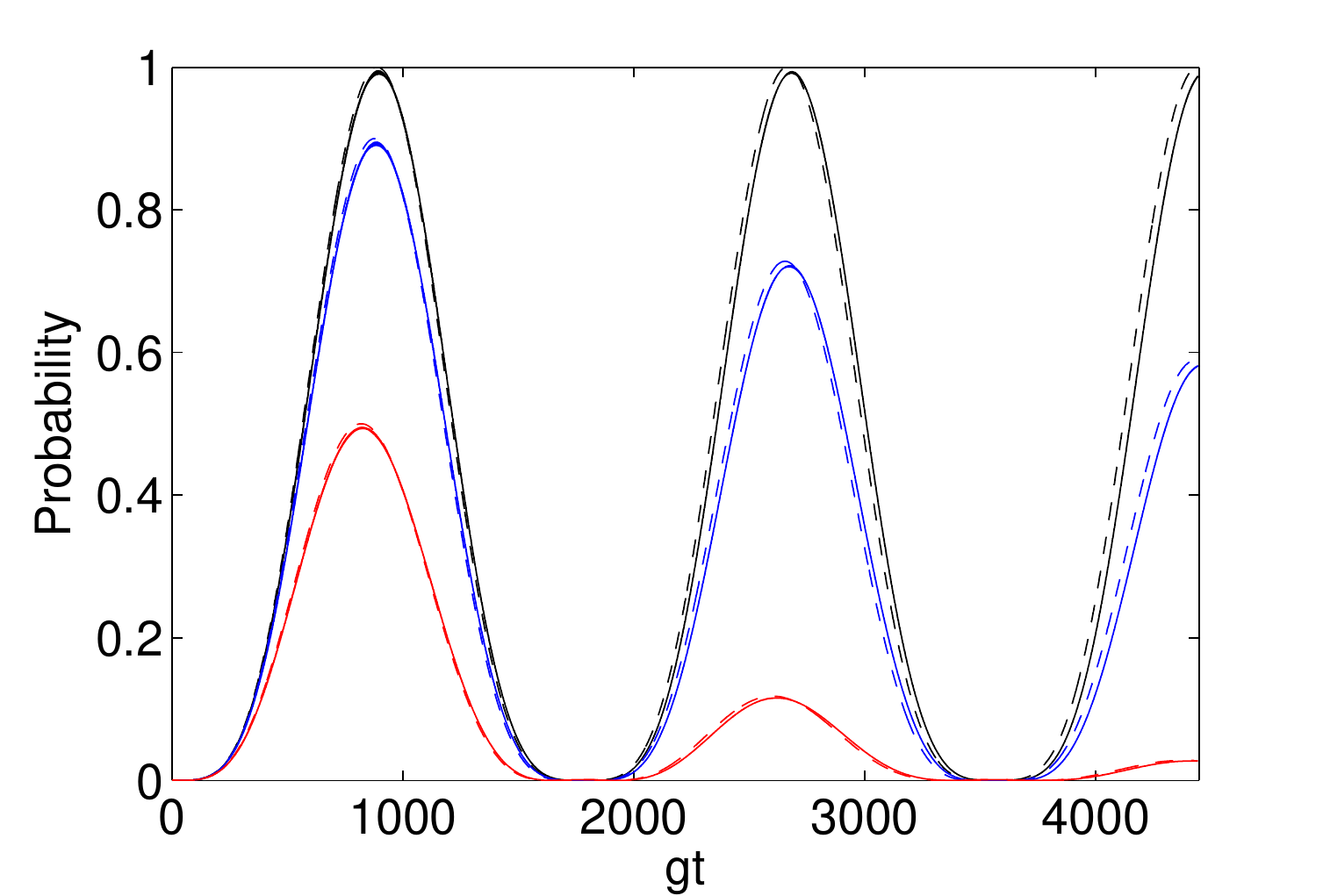}}
  \subfigure[]{\includegraphics[width=.32\textwidth]{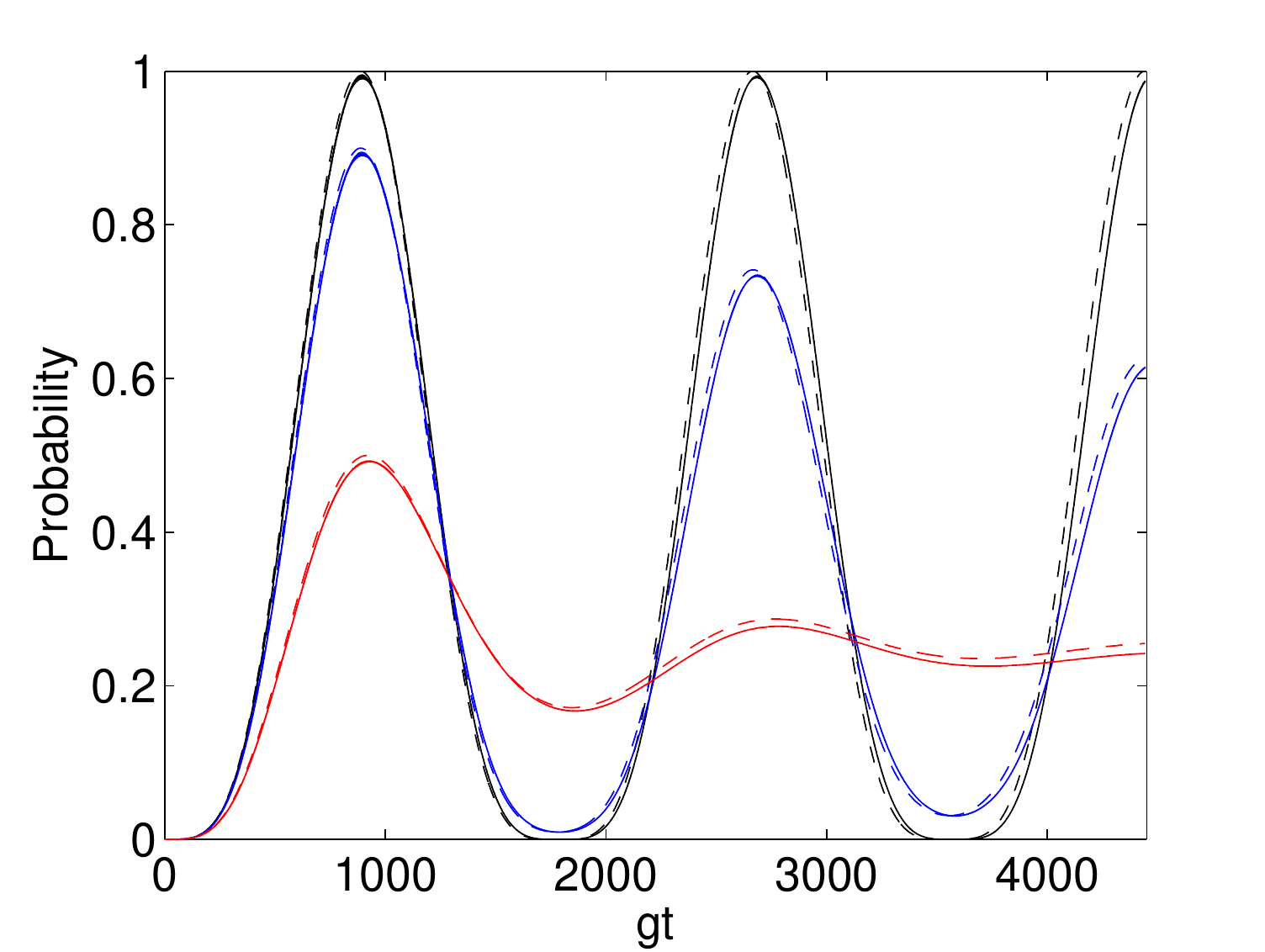}}
  \subfigure[]{\includegraphics[width=.38\textwidth]{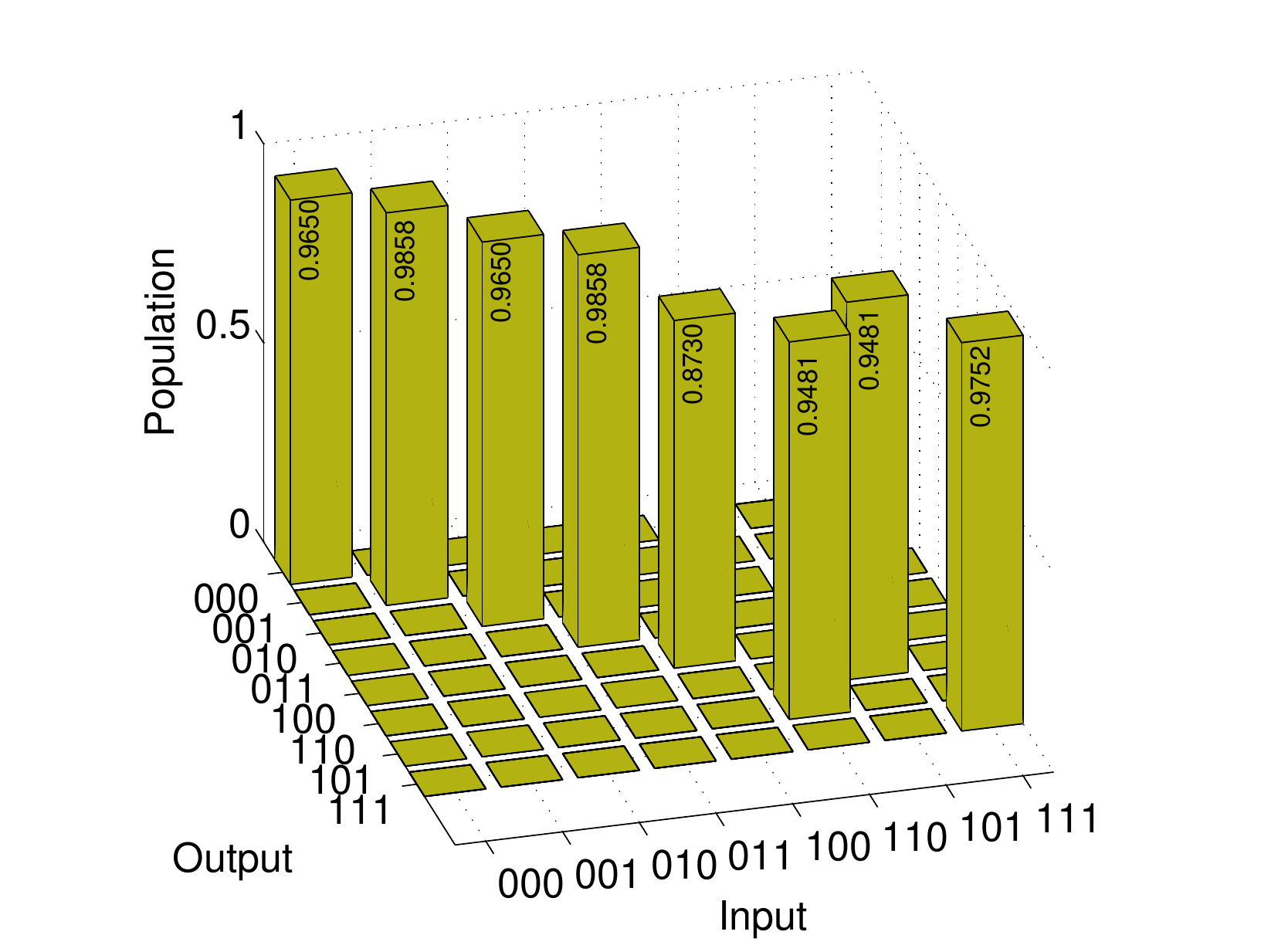}}}
  
  \caption{In (a) and (b) the probability of the qubit state $|a\ 110\rangle$
    in the model (1001001) with non-vanishing photonic and atomic
    decays. Solid and dashed curves show the numerical and theoretical
    solutions, respectively. The coupling constants $g_{i}$ ($i=1,2,3,5,6$)
    all are set to $g$, and the detunings $\Delta_{1,2,4,5}=\Delta$ with 
    $\Delta=20 g$. The effective couplings of the truncated system
    $\Gee{(1)}{\rm{eff}}$ and $\Gee{(2)}{\rm{eff}}$ are given by
    Eq.~(\ref{eq:Fredkin-g-eff}), and the detunings $\Delta_3$ and $\Delta_6$
    are defined by the resonance conditions in
    Eq.~(\ref{eq:Fredkin-Delta-eff}). The black, blue, and red 
    lines respectively represent (at the interaction time $gt_{int}$) the
    system probability $P\sim (100, 90, 50) \%P_0$ with a maximum probability
    $P_0 \sim 0.9950$. The three curves corresponding to the values of: (a)
    the photonic decay rates $\kappa \sim (0,\ 0.0174,\
    0.1186)\bar{g}/\sqrt{2}$ with $\Gamma \mapsto 0$, and (b) the atomic decay
    rates $\Gamma \sim (0,\ 0.0976,\ 0.764)\bar{g}/\sqrt{2}$ with $\kappa
    \mapsto 0$. (c) a truth table of the numerically simulated Fredkin gate
    with $\Delta=10\ g$ in the presence of decoherence processes. Parameters:
    the coupling constant is approximately $g/2\pi=50$ kHz, $\Gamma/g \sim
    10^{-4}$, and $\kappa/g \sim
    2.5\times10^{-5}$.} \label{fig:DampedFredkinGate} 
\end{figure}
\end{widetext}
In the absence of $\Gamma$ it is shown by
Eqs.~(\ref{eq:FastIswapPopulationLoss}, \ref{eq:FastFredkinPopulationLoss})
that the Fredkin gate is more sensitive to cavity field decay rate, and the
i\textsc{swap} and the Fredkin gates in the previous three-level configurations have
the same sensitivity to the spontaneous emissions when no photonic decay is
considered.

\subsection{Single qubit gates} 
\label{SingleQubitGates}

\begin{figure}\centering
  
  \includegraphics[width=.3\textwidth]{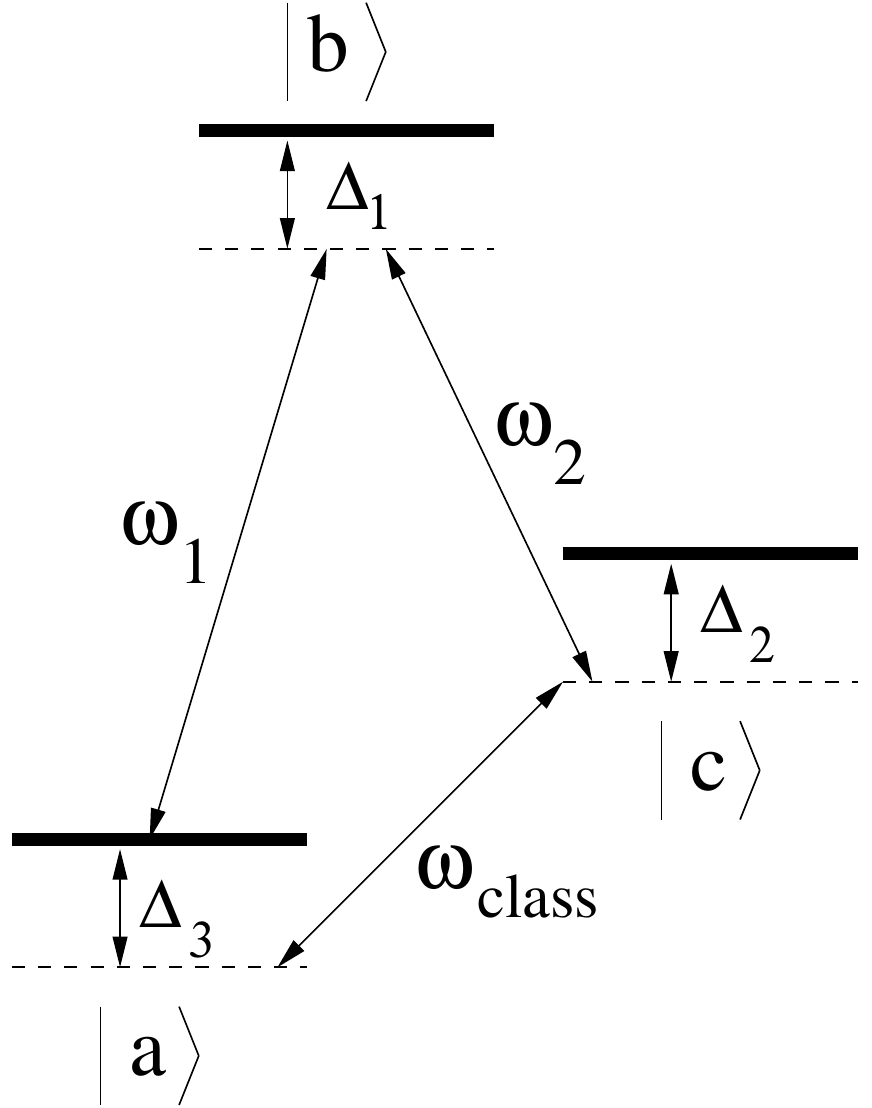}
  
  \caption{The model for a single-qubit Pauli-X gate. A three-level atom with
    $\Lambda$ configuration of levels interacting with the optical cavity
    modes $\omega_1$ and $\omega_2$ on the transitions $|a\rangle
    \leftrightarrow |b\rangle$ and $|b\rangle \leftrightarrow |c\rangle$. The
    transition $|c\rangle \leftrightarrow |a\rangle$ is coupled by the
    classical field $\omega_{class}$.}
\label{fig:QubitRotations}
\end{figure}
 We turn briefly to single qubit rotations which are required to make
 a universal set of gates. Within the dual-mode scheme a
 $\Lambda$-atom-mode system, such as seen in Fig.~\ref{fig:Larson}
 cannot be used because for an arbitrary rotation of the mode
 excitation between the cavity modes the atomic state changes as
 well. For that reason we add a classical field to allow a return to
 the original atomic state $\k a$ (see
 Fig.~\ref{fig:QubitRotations}). Otherwise the scheme is similar to
 Fig.~\ref{fig:Larson} in that we have a lambda atom which has two
 transitions  coupled to two cavity modes that make up a qubit. We will
 adiabatically eliminate levels $\ket{b}$ and $\ket{c}$ from the interaction 
 under the conditions $\Delta_1, \, \Delta_2  \gg  \Gee{ab}{1}, \,
 \Gee{bc}{2}, \,\Omega/2, \, \Delta_3$ (see appendix~\ref{NotGate}) to find
 the effective detuning
\begin{equation}\label{xRot-Detuning}
\eff{\Delta} =  \Delta_3 + \frac{(\Gee{ab}{1})^2}{\Delta_1}\ ,
\end{equation}
and the effective coupling 
\begin{equation}\label{xRot-g}
\eff{g} = \frac{\Gee{ab}{1} \Gee{bc}{2}\Omega}{2 \Delta_1\Delta_2}\ .
\end{equation}
We can then ensure a qubit rotation in the form
\begin{equation}
\mathrm{\hat R}_x(g_\mathrm{eff}t) = \cos(g_\mathrm{eff}t/2) \rm{\hat I}
            -\mathrm{i}\sin(g_\mathrm{eff}t/2) \ \hat\sigma_x\ .
\end{equation}

The Pauli Z gate can be easily realized by our qubits, too. Generally
speaking, the atom-cavity interaction in the Jaynes-Cummings model shows that 
for an atom in the ground state $|g\rangle$ interacting with a single mode
having $n$ photons:
\begin{equation*}
 |g,n\rangle \mapsto \cos(g\sqrt{n}t)|g,n\rangle -\rm{i} \sin(g \sqrt{n}t)
 |e,n-1\rangle\ .
\end{equation*}
This is the case when the frequencies of the atomic transitions and the mode
are equally matched (i.e. the resonance case $\omega_{eg}=\omega$). In the
case of very large detuning ($\Delta \gg g$), on the other hand, the system
remains in its initial state and a phase shift can be produced as
\begin{eqnarray}
|g,n\rangle \mapsto e^{\rm{i}\Phi(n)}|g,n\rangle \ ,
\end{eqnarray}
with $\Phi(n)$ can be expressed as \cite{Walther5}
\begin{eqnarray}
\Phi(n)=\frac{\Delta}{2v} \int_0^L
dz\bigg[\sqrt{1+n\bigg(\frac{g(z)}{\Delta/2}\bigg)^2}-1 \bigg] \ ,
\end{eqnarray}
where $v$ is the velocity of the atom passing through a cavity, $L$ is the
cavity length, and $g(z)$ is the coupling constant which in our case is
independent of z. By setting $n=0$ nothing happens, but with a cavity being
initially in the number state $|1\rangle$ (i.e.\ there is $n=1$ photon) a
set of phase gates can be realised and the rotation operator $R_{z}$ can be
produced. In the case of two modes inside the cavity interacting with
an atom in the ground state $|a\rangle$, such a case in the dual-rail qubits
$|a\ 10\rangle$ and $|a\ 01\rangle$, the previous argument can be followed to
introduce a phase shift. That is, we can set a large detuning
between the atom and, say the first mode, and set a very high detuning between
the atom and the second mode. In this case, if the excitation is in the first
mode, one finds $|a\ 10\rangle \mapsto e^{\rm{i}g^2 t/\Delta}|a\ 10\rangle$;
otherwise, $|a\ 01\rangle \mapsto |a\ 01\rangle$.

Different ways can be considered to add a global phase
$\eta t$ to the two- or three-qubit gates already discussed in the previous
sections. For instance, the single-qubit phase gate above can be employed for
this purpose. That is, in the case of the previous i\textsc{swap} gate and after
producing the transformations in Sec.~\ref{sec:FourModesVariants}, an atom
initially in the ground state $|a\rangle$ sent to the first two-mode cavity can
introduce a phase to the logical state $|a\ 10\rangle$ when the atom is
detuned from the first mode $\hat{n}_1$ of a qubit, and far detuned from the
mode $\hat{n}_2$. Then, another atom in $|a\rangle$ interacting with the
second two-mode cavity, where the atom is detuned from $\hat{n}_4$ and far
detuned from $\hat{n}_3$, can add a phase to the logic $|a\ 01\rangle$. 


\section{Conclusion}
\label{sec:Conclude}

A multiphoton resonance can be a very useful technique for applications
in quantum information processing as it involves a process conditional on
the presence of various photons. Here we use the theory of
multiphoton resonance with a multilevel multiphoton Jaynes-Cummings model. 
Information is stored in photonic qubits and we produce a set of practical
one-qubit, two-qubit, and even three-qubit gates. This works because CQED
offers a high non-linearity with low losses. 

Up to date, the strong interaction between a multilevel atom with a
multimode field (such an interaction proposed in our scheme)
remains an experimental challenge. In fact, with the remarkable progress in
nanotechnology, this kind of interaction might be possible in near
future \cite{Zubairy3}. It is reported in \cite{Benson} that a transfer of
energy between two individual nanoparticles strongly coupled to high-Q
whispering-gallery modes in a microsphere resonator is experimentally
achieved. This achievement gives a great hope of finding experiments that
proceed an interaction between a single multimode cavity interacting with a
multilevel atom in the limit of strong coupling. 


MMA would like to acknowledge support from King Khalid University (KKU).
MSE thanks the Japanese Society for the Promotion of Science.  BMG
acknowledges the support of the Leverhulme Trust and thanks Michael Hartmann
and Bruce W.\ Shore for comments.


\appendix

\section{Adiabatic elimination}
\label{sec:TheoryAdiabaticElimination}

Generally, the time evolution of a quantum system is governed by the Schr\"odinger equation:
\begin{equation}
   i \frac{\partial}{\partial t}  \mathbf{c}(t) = \mathbf{H}  \, \mathbf{c}(t)
\end{equation}
Applying a Laplace transform shows that 
\begin{equation}
  i( s   \overline{\mathbf{c}}(s) - \mathbf{c}(0) ) = \mathbf{H} \,
  \overline{\mathbf{c}}(s)\ .
\end{equation}
An approximate solution in Laplace space, therefore, can be expressed as 
\begin{equation}
  \overline{\mathbf{c}}(s) = ( s \mathbf{I} + i \mathbf{H} )^{-1}
  \mathbf{c}(0)\ .
\end{equation}

By assuming that the Hamiltonian $\mathbf{H}$ to be represented by a $2\times 2$ matrix where the
states of interest are included in $\mathbf{W}_0$ and the states eliminated under 
certain conditions are represented by the matrix $\mathbf{A}$, $\mathbf{H}$ can be defined as\\ 
$\mathbf{H} = \begin{bmatrix} 
      \mathbf{W_0} & \mathbf{B}\\
      \mathbf{B^\dagger} & \mathbf{A}\\
\end{bmatrix}$.
\begin{widetext}
Then, the inverse of the square matrix $(s \mathbf{I} + i \mathbf{H})$ reads 
\begin{eqnarray}
( s \mathbf{I} + i \mathbf{H} )^{-1}& = \left[
\begin{array}{cc}& -i\mathbf{X}\mathbf{B} ( s + i\mathbf{A})^{-1} \\
      -i( s + i\mathbf{A})^{-1}\mathbf{B}^\dagger\mathbf{X} &
      ( s + i\mathbf{A})^{-1} - ( s + i\mathbf{A})^{-1} \mathbf{B}^\dagger\mathbf{X}\mathbf{B}( s + i\mathbf{A})^{-1}
\end{array} \right]\ ,\label{eq:BarryApproach1}
\end{eqnarray}
where $\mathbf{X}=[ s + i\mathbf{W}_0 + \mathbf{B}( s + i\mathbf{A})^{-1} \mathbf{B}^\dagger ]^{-1}
$.
\end{widetext}

Equation~(\ref{eq:BarryApproach1}) can be further simplified by introducing
the approximation that the eigenvalues of $\mathbf{A}$ are much larger in magnitude
than the eigenvalues of $\mathbf{W}_0$ \cite{Shore}. One then finds that 
\begin{equation*}
    ( s \mathbf{I} + i \mathbf{H} )^{-1} 
  \sim
  \begin{bmatrix} 
      [s + i(   \mathbf{W}_0 -   \mathbf{B}\mathbf{A}^{-1} \mathbf{B}^\dagger)     ]^{-1}  &
      {\cal O}(1/A)\\
      {\cal O}(1/A)&
      {\cal O}(1/A)\\
  \end{bmatrix}
\end{equation*}
The subsystem containing only the states of interest can be, therefore,
described by the effective Hamiltonian 

$\mathbf{H}_{\rm{eff}}=\mathbf{W}_0 - \mathbf{B}\mathbf{A}^{-1}
\mathbf{B}^\dagger $.


\section{The i\textsc{swap} Gate}\label{app:TheIswapGate}
\subsection{A two-level approximation}\label{app:TheIswapGate1}
In Sec.~\ref{sec:MultiphotonLogic}, we have seen that in the interaction
picture and with $|a\ 1010\rangle$ to be the initial state the Hamiltonian of
the atom-field system is: 
\begin{eqnarray}
H^{\prime}=\left(\begin{array}{ccccc}0&\Gee{ab}{1}&0&0&0\\
                                    \Gee{ab}{1}&\Delta_1&\Gee{bc}{2}&0&0\\
                                    0&\Gee{bc}{2}&\Delta_2&\Gee{cd}{3}&0\\
                                    0&0&\Gee{cd}{3}&\Delta_3&\Gee{da}{4}\\
                                    0&0&0&\Gee{da}{4}&\Delta_4 \end{array}\right)\
                                ,\label{eq:iSWAP_int_Ham-dup} 
\end{eqnarray}
where the system detunings $\Delta_i$ ($i=1,2,3,4$) can be defined as
\begin{eqnarray}
\nonumber
\Delta_1&=&(\omega_{ba}-\omega_1)\ ,\\
\Delta_2&=&(\omega_{ba}-\omega_1)-(\omega_{bc}-\omega_2)\ ,\label{eq:iSWAP detunings}\\
\nonumber
\Delta_3&=&(\omega_{ba}-\omega_1)-(\omega_{bc}-\omega_2)+(\omega_{dc}-\omega_3)\ ,\\
\nonumber
\Delta_4&=&(\omega_{ba}-\omega_1)-(\omega_{bc}-\omega_2)+(\omega_{dc}-\omega_3)-(\omega_{da}-\omega_4)\ . 
\end{eqnarray}
By following Shore's method above, the basis states given by $|\Psi(t)\rangle$ in 
Eq.~(\ref{eq:fourmodebasis}) can be divided into a couple of
subsystems $\mathbb{P}|\Psi(t)\rangle$ and $\mathbb{Q}|\Psi(t)\rangle$, where
$\mathbb{P}$ and $\mathbb{Q}$ are 
orthogonal projection operators and $\mathbb{P}+\mathbb{Q}=1$. Assuming
$\mathbb{P}$ consists of the states 
$|a10\rangle$ and $|a01\rangle$, one finds that the operators
$H_0=\mathbb{P}H\mathbb{P}$, 
$A=\mathbb{Q}H\mathbb{Q}$, and $B=\mathbb{P}H\mathbb{Q}$ can be expressed, in
the matrix formalism, as 
\begin{eqnarray}
\nonumber 
H_0= \left[\begin{array}{cc}0&0\\0&\Delta_4 \end{array}\right]\ ,\ 
B&=& \left[\begin{array}{ccc}\Gee{ab}{1}&0&0\\0&0&\Gee{ad}{4}\end{array}\right]\ ,\ \\
A&=& \left[\begin{array}{ccc}\Delta_1&\Gee{bc}{2}&0\\\Gee{bc}{2}&\Delta_2&\Gee{cd}{3}\\0&
\Gee{cd}{3}&\Delta_3 \end{array}\right]\ .
\end{eqnarray}
An effective two-level Hamiltonian $H_{\rm{eff}}$ can be constructed by
$H_{\rm{eff}}=H_0-B\ A^{-1}\ B^{\dag}$. The effective coupling is
found to be 
\begin{align}\label{eq:geff-ISWAP-slow-dup}
\nonumber
\eff{g}&=-\frac{\Gee{ab}{1}\Gee{bc}{2}\Gee{cd}{3}\Gee{da}{4}}{\Delta_1\Delta_2\Delta_3-
\Delta_3(\Gee{bc}{2})^2-\Delta_1(\Gee{cd}{3})^2}\ ,\\
&\approx-\frac{\Gee{ab}{1}\Gee{bc}{2}\Gee{cd}{3}\Gee{da}{4}}{\Delta_1\Delta_2\Delta_3}\ ,
\end{align}
and the effective detuning of the two-level system is 
\begin{eqnarray} \label{eq:Delta-ISWAP-slow-dup}
\nonumber
&&\Delta_{\rm{eff}}=\\
\nonumber
&&\Delta_4+\frac{(\Gee{ab}{1})^2(\Delta_2\Delta_3-(\Gee{cd}{3})^2)-(\Gee{da}{4})^2(\Delta_1\Delta_2-(\Gee{bc}{2})^2)} 
{\Delta_1\Delta_2\Delta_3-\Delta_3(\Gee{bc}{2})^2-\Delta_1(\Gee{cd}{3})^2}\ ,\\
&&\approx
\Delta_4+\frac{(\Gee{ab}{1})^2}{\Delta_1}-\frac{(\Gee{da}{4})^2}{\Delta_3} \ .
\end{eqnarray}

\subsection{A three-level system}\label{app:TheIswapGate2}
In the case of $\mathbb{Q}=|c,0110\rangle \langle c,0110|+|d,0100\rangle
\langle d,0100|$, the operators $H_0$, $B$, and $A$ can be given as  
\begin{eqnarray}\nonumber
H_{0}=\left[\begin{array}{ccc} 0  &\Gee{ab}{1}     &0\\
                               \Gee{ab}{1}&\Delta_1&0\\
                               0  &0       &\Delta_4
           \end{array}\right],\
B&=&\left[\begin{array}{cc}    0  &0  \\
                               \Gee{bc}{2}&0  \\
                               0  &\Gee{da}{4} 
        \end{array}\right],\ \\
A&=&\left[\begin{array}{cc}\Delta_2&\Gee{cd}{3}\\
                           \Gee{cd}{3}     &\Delta_3\end{array} \right].\
\end{eqnarray}
In the space $\{|a1010\rangle,|b0010\rangle,|a0101\rangle\}$, the effective
Hamiltonian $H_\mathrm{eff}$ can be constructed as 
\begin{equation}
\label{3levelmatrix}
H_{\mathrm{eff}}=\begin{bmatrix}
0 & \Gee{eff}{(1)} & 0\\
\Gee{eff}{(1)} & \Delta_1^{\mathrm{eff}} & \Gee{eff}{(2)}\\
0 & \Gee{eff}{(2)} & \Delta_2^{\mathrm{eff}}
\end{bmatrix}\ ,
\end{equation}
and then the corresponding effective couplings and detunings can be expressed
as 
\begin{eqnarray}
g_{\rm{eff}}^{(1)}=g_1^{\rm{ab}} ,\ 
g_{\rm{eff}}^{(2)}=\frac{g^{\rm{bc}}_2g^{\rm{cd}}_3g^{\rm{da}}_4}{(\Delta_2\Delta_3-(\Gee{cd}{3})^2)} 
\approx\frac{g^{\rm{bc}}_2g^{\rm{cd}}_3g^{\rm{da}}_4}{\Delta_2\Delta_3} \
, \label{eq:effective coupling in 11001-dup} 
\end{eqnarray}
and the effective detunings
\begin{eqnarray}
\nonumber
\Delta_{\rm{eff}}^{(1)}&=&\Delta_1-\frac{(\Gee{bc}{2})^2\Delta_3}{(\Delta_2\Delta_3-(\Gee{cd}{3})^2)}
\approx\Delta_1-\frac{(\Gee{bc}{2})^2}{\Delta_2}\ ,\\
\Delta_{\rm{eff}}^{(2)}&=&\Delta_4-\frac{(\Gee{da}{4})^2\Delta_2}{(\Delta_2\Delta_3-(\Gee{cd}{3})^2)}
\approx\Delta_4-\frac{(\Gee{da}{4})^2}{\Delta_3}\ .
\label{eq:resonance conditions in 11001-dup} 
\end{eqnarray}

\section{A Fast Fredkin Gate}\label{sec:FastFredkinGate}

\begin{figure}\centering
  
  \includegraphics[width=\bmgstdfigwidth]{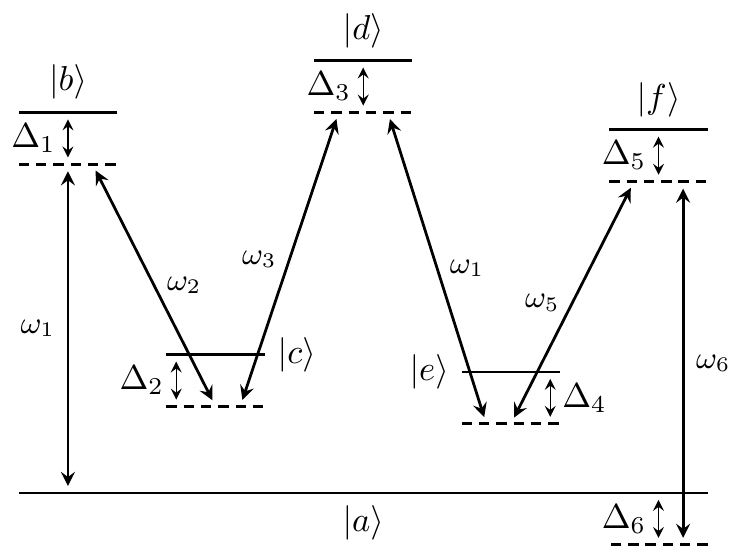}
  
  \caption{Energy level scheme for a Fredkin gate. Note the repeated mode
    $\omega_1$. Operation requires at least $\Delta_6\sim 0$.}
\label{fig:FredkinLevels}
\end{figure}

The Fredkin gate, as a three-qubit gate, can be realized in our scheme as 
mentioned in Sec.~\ref{sec:ThreeAndOneQubitGates}. Considering
Fig.~\ref{fig:FredkinLevels}, this system is governed by the Hamiltonian $H$ 
\begin{eqnarray}
\nonumber 
H&=&\sum \limits_i \omega_i \sigma_{ii}+\sum \limits_j \omega_j
a^{\dag}_ja_j \\
&+&[g_1\ \sigma_{ba}\ a_1+g_2\ \sigma_{cb}\ a^{\dag}_2\\
\nonumber
&+&g_3\ \sigma_{dc}\ a_3+g_1\ \sigma_{ed}\ a^{\dag}_1+g_5\ \sigma_{fe}\ a_5
+g_6 \sigma_{af}\ a^{\dag}_6+ \rm{H.c.}]\ ,\label{eq:general Ham in Fredkin}
\end{eqnarray}
where $(i=a,b,c,d,e,f)$ and $(j=1,2,3,5,6)$.
Beginning with the initial state of the atom-field system $|\Psi(0)\rangle=|a\ 
10,01,10\rangle \equiv|a\ 101\rangle$, the time evolution of $|\Psi(0)\rangle$
can be a superposition 
\begin{eqnarray}
\nonumber
|\Psi(t)\rangle
&=&c_1(t)\ |a\ 10,01,10\rangle +c_2(t)\ |b\ 00,01,10\rangle\\
\nonumber
&+&c_3(t)\ |c\ 00,11,10\rangle +c_4(t)\ |d\ 00,10,10\rangle\\
\nonumber
&+&c_5(t)\ |e\ 10,10,10\rangle +c_6(t)\ |f\ 10,10,00\rangle\\
\nonumber
&+&c_7(t)\ |a\ 10,10,01\rangle +c_8(t)\ |b\ 00,10,01\rangle\\
&+&c_9(t)\ |c\ 00,20,01\rangle\ .\label{eq:state of qubit a101} 
\end{eqnarray}
As shown above, the system does not terminate in the atomic state $|a\rangle$,
and this system has the over-shot states $|b\ 00,10,01\rangle$ and $|c\
00,20,01\rangle$. To avoid these states to be populated, we always assume the
detunings $\Delta_1$ and $\Delta_2$ to be very large. The definitions of the
detunings in Eq.~(\ref{eq:iSWAP detunings}) can be easily employed to define
the system detunings $\Delta_{i}$ (with $i=1,2,...,6$) in the Fredkin gate. 

An effective three-level behaviour can be analysed by allowing the
$\mathbb{P}$-space to include $|a,101\rangle$, $|d\rangle$, and 
$|a,110\rangle$. The states $|b,00,01,10\rangle, |c,00,11,10\rangle,
|e,10,10,10\rangle, |f,10,10,00\rangle,\\|b,00,10,01\rangle$, and
$|c,00,20,01\rangle$  must be off-resonant so that they remain
unpopulated. The required operators for the effective Hamiltonian 
$H_{\rm{eff}}=H_0-BA^{-1}B^{\dag}$ can  be expressed as 

\begin{eqnarray}\nonumber
H_0&=&\left[\begin{array}{ccc}
0&0       &0       \\
0&\Delta_3&0       \\
0&0       &\Delta_6\end{array}\right],\  
B=\left[\begin{array}{cccccc}
\Gee{ab}{1}&0          &0          &0          &0          &0\\
0          &\Gee{cd}{3}&\Gee{de}{1}&0          &0          &0\\
0          &0          &0          &\Gee{af}{6}&\Gee{ab}{1}&0\end{array}\right], \\
A&=&\left[\begin{array}{cccccc}
\Delta_1    & \Gee{bc}{2} & 0           & 0           & 0 & 0\\ 
\Gee{bc}{2} & \Delta_2    & 0           & 0           & 0 & 0\\
0           & 0           & \Delta_4    & \Gee{ef}{5} & 0 & 0\\
0           & 0           & \Gee{ef}{5} & \Delta_5    & 0 & 0\\
0           & 0           & 0& 0& \Delta_6+\Delta_1&\Gee{bc}{2}\sqrt{2}\\ 
0 & 0 & 0 & 0 & \Gee{bc}{2}\sqrt{2} & \Delta_6+\Delta_2\end{array}\right]\ .
\end{eqnarray}

Within states $\ket{10,01,10,a}$, $\ket{00,10,10,d}$ and $\ket{10,10,01,a}$,
the effective Hamiltonian $H_{\rm{eff}}$ can be given by
Eq.~(\ref{3levelmatrix}) where
\begin{align*}
\Gee{eff}{(1)} &= \frac{\Gee{ab}{1}\Gee{bc}{2}\Gee{cd}{3}}{\Delta_1\Delta_2},&
\Gee{eff}{(2)} & = \frac{\Gee{de}{1}\Gee{ef}{5}\Gee{fa}{6}}{\Delta_4\Delta_5}
\,,
\end{align*}
and
\begin{equation*}
\begin{split}
&\Delta_1^{\mathrm{eff}}\approx\Delta_3 +
\frac{\left(\Gee{ab}{1}\right)^2}{\Delta_1} -
\frac{\left(\Gee{cd}{3}\right)^2}{\Delta_2}-\frac{\left(\Gee{de}{1}\right)^2}{\Delta_4}\,,\\
&\Delta_2^{\mathrm{eff}}\approx\Delta_6 - \frac{\left(\Gee{af}{6}\right)^2}{\Delta_5}\,.
\end{split}
\end{equation*}

\section{The \textsc{not} gate}\label{NotGate}
The realisation of the single-qubit \textsc{not} gate is possible in our
scheme. Considering the model in Fig.~\ref{fig:QubitRotations}, we 
assume a de-excited three-level atom in the $\Lambda$ configuration interacts
with a dual-rail photonic qubit $|10\rangle$ or $|01\rangle$. The initial
state, therefore, can be either the logic $|a\ 10\rangle$ or $|a\ 01\rangle$. Then, the
atom interacts with a classical field on the transition $|c\rangle \mapsto
|a\rangle$. The corresponding Hamiltonian describing all such interactions,
i.e. the cavity-atom interaction plus the classical field-atom interaction,
can be defined as 
\begin{eqnarray}
H&=&\omega_a\sigma_{aa}+\omega_b\sigma_{bb}+\omega_c\sigma_{cc}+\omega_1a^{\dag}_1a_1+\omega_2a^{\dag}_2a_2\\ 
\nonumber&+&
\bigg[\Gee{ab}{1}a^{\dag}_1\sigma_{ab}+\Gee{bc}{2}\sigma_{bc}a_2+(\Omega/2)e^{\rm{i}\omega_3t}\sigma_{ac}+
\rm{H}.\rm{C}.\bigg].  
\end{eqnarray}
Given the system in the initial state $|\Psi(0)\rangle=|a\ 10\rangle$, this state evolves into
a superposition
\begin{eqnarray}
\nonumber
|\Psi(t)\rangle &=&c_{c10}(t)\ |c\ 10\rangle+ c_{a10}(t)\ |a\
10\rangle+c_{b00}(t)\ |b\ 00\rangle\\ 
&+&c_{c01}(t)\ |c\ 01\rangle+c_{a01}(t)\ |a\
01\rangle \ .
\end{eqnarray}
Then, by using the Schr\"odinger equation $\frac{\partial}{\partial
t}|\Psi(t)\rangle=-\rm{i}H|\Psi(t)\rangle$ a set of amplitude equations, within
the rotating wave approximation, can be obtained. To transform these amplitude
equations to the frame rotating with the frequencies of the optical fields
$\omega_1$, $\omega_2$, and $\omega_{class}$, we introduce the
transformation (note that the initial state $|a\ 10\rangle$ is set as a zero
point energy) 
\begin{eqnarray}
\nonumber
c_{c10}(t)&=& c_{c10}^{\prime}(t)\ e^{-\rm{i}(\omega_a+\omega_{class})t}\ e^{-\rm{i}\omega_1 t};\\ 
\nonumber
c_{a10}(t)&=& c_{a10}^{\prime}(t)\ e^{-\rm{i}\omega_a t}\ e^{-\rm{i}\omega_1 t};\\
c_{b00}(t)&=& c_{b00}^{\prime}(t)\ e^{-\rm{i}(\omega_a+\omega_1)t};\\
\nonumber
c_{c01}(t)&=& c_{c01}^{\prime}(t)\ e^{-\rm{i}(\omega_a+\omega_1-\omega_2)t}\ e^{-\rm{i}\omega_2};\\
\nonumber
c_{a01}(t)&=& c_{a01}^{\prime}(t)\ e^{-\rm{i}(\omega_a+\omega_1-\omega_2-\omega_{class})t}\ e^{-\rm{i}\omega_2 t}\ .
\end{eqnarray}
The RWA Hamiltonian, then, can be re-expressed as 
\begin{eqnarray}
H^{\prime}=\left[\begin{array}{ccccc}
(\Delta_2-\Delta_3)&\Omega/2&0       &0       &0\\
           \Omega/2&0       &\Gee{ab}{1}     &0       &0\\
0                  &\Gee{ab}{1}     &\Delta_1&\Gee{bc}{2}     &0\\
0                  &0       &\Gee{bc}{2}     &\Delta_2&\Omega/2\\
0                  &0       &0       &\Omega/2&\Delta_3
\end{array}\right]\ ,
\end{eqnarray}
where $H^{\prime}$ acts in the basis $\{|c\ 10\rangle\, |a\ 10\rangle, |b\
00\rangle, |c\ 01\rangle, |a\ 01\rangle \}$. Now the basis states other than
$|a\ 10\rangle$ and $|a\ 01\rangle$ can be adiabatically eliminated by
recalling Shore's method. That is, we allow large values for $\Delta_1,\
\Delta_2$. In other words, we assume the sates $|a\ 10\rangle$ and $|a\
01\rangle$ to be spanned by the space $\mathbb{P}$ and the remaining states to
be set in the space of $\mathbb{Q}$. The resultant operators, then, can be
given as
\begin{eqnarray}
\nonumber
H_0=\left[\begin{array}{cc}0&0\\0&\Delta_3 \end{array}\right]\ ;
B&=&\left[\begin{array}{ccc}\Omega/2&\Gee{ab}{1}&0\\0&0&\Omega/2\end{array}\right]\ ;\\
A&=&\left[\begin{array}{ccc}(\Delta_2-\Delta_3)&0&0\\0&\Delta_1&\Gee{bc}{2}\\0&\Gee{bc}{2}&\Delta_2 \end{array}\right]. 
\end{eqnarray}
The corresponding effective detuning with
$\Delta_{1,2} \gg g_{1,2},\Omega/2,\Delta_3$ is 
\begin{eqnarray}
\nonumber
H_{\rm{eff}}&=&\Delta_3-\frac{(\Omega/2)^2\Delta_1}{(\Delta_1\Delta_2-(\Gee{bc}{2})^2)}+
\frac{(\Omega/2)^2}{(\Delta_2-\Delta_3)}\\ 
\nonumber
&+&\frac{(\Gee{ab}{1})^2\Delta_1}{(\Delta_1\Delta_2-(\Gee{bc}{2})^2)}\\
&\approx&\Delta_3+\frac{(\Gee{ab}{1})^2}{\Delta_2}\ ,
\end{eqnarray}
and the effective coupling strength is
\begin{eqnarray}
g_{\rm{eff}}=\frac{(\Omega/2)\Gee{ab}{1}\Gee{bc}{2}}{(\Delta_1\Delta_2-(\Gee{bc}{2})^2)}\approx
\frac{\Omega\ \Gee{ab}{1}\Gee{bc}{2}}{2\Delta_1\Delta_2}\ .
\end{eqnarray}
At the resonance condition, the time evolution of the initial state $|a\
10\rangle$ or $|a\ 01\rangle$ can be given by
Eq.~(\ref{eq:iSWAP-slow-2Eqns}), and with an appropriate interaction time
$g_{\rm{eff}}t_{int}$ and a global phase the Pauli $X$ gate can be easily
realized, and the exponential of the \textsc{not} gate is nothing but the rotation
operator $\rm{R}_{x}(g_{\rm{eff}}t)$.

\bibliography{BarryMoteb}

\end{document}